\documentclass[journal,a4paper,10pt]{IEEEtran}

\usepackage{citesort,amsmath,upref,amssymb,color,bm}
\usepackage{epsfig,graphicx,rotating}

\allowdisplaybreaks

\newcommand{\E}[1]{{\rm E} {}}

\DeclareMathOperator{\sinc}{sinc}               

\newcommand{\ist}{\hspace*{.3mm}}
\newcommand{\rmv}{\hspace*{-.3mm}}
\newcommand{\remark}[1]{}
\newcommand{\isum}[1]{\sum_{#1 = -\infty}^{\infty}} 
\newcommand{\esum}[2]{\sum_{#1}^{#2}} 

\newcommand{\T}{T}
\newcommand{\ppsi}{\psi}

\def\C{\mathbb{C}}
\def\Z{\mathbb{Z}}

\def\M{M} 

\definecolor{grau}{gray}{0.6}

\newcommand{\eq}{\,=\,}
\renewcommand{\jmath}{j}

\newcommand{\be}{\begin{equation}}
\newcommand{\ee}{\end{equation}}

\newcommand{\aalpha}{\bm{\alpha}\hspace*{-2.7mm}\bm{\alpha}}
\newcommand{\bbeta}{\bm{\beta}\hspace*{-2.55mm}\bm{\beta}}
\newcommand{\ttheta}{\bm{\theta}\hspace*{-2.1mm}\bm{\theta}}
\newcommand{\vvartheta}{\bm{\vartheta}\hspace*{-2.4mm}\bm{\vartheta}}
\newcommand{\iter}{r}
\newcommand{\sleb}{\lfloor \nu_ {\max}T_{\rmv{\rm s}} N_{r}\rfloor}

\begin{document}

\title{Compressive Estimation of Doubly Selective\\
Channels in Multicarrier Systems: Leakage\\
Effects and Sparsity-Enhancing Processing\thanks{ Manuscript
received February 27, 2009; revised October 17, 2009. Current
version published March 17, 2010. This work was supported by WWTF
grants MOHAWI (MA 44) and SPORTS (MA 07-004) and by FWF Grants
``Signal and Information Representation'' (S10602-N13) and
``Statistical Inference'' (S10603-N13) within the National
Research Network SISE. The work of H. Rauhut was supported by the
Hausdorff Center for Mathematics, University of Bonn. Parts of
this work have been previously published at IEEE ICASSP 2008, Las
Vegas, NV, March--April 2008 and at EUSIPCO 2008, Lausanne,
Switzerland, Aug. 2008. The associate editor coordinating the
review of this manuscript and approving it for publication was Dr.
Yonina Eldar.\newline\hspace*{2mm} G. Taub\"{o}ck and F. Hlawatsch are
with the Institute of Communications and Radio-Frequency
Engineering, Vienna University of Technology, A-1040 Vienna,
Austria (e-mail: gtauboec@nt.tuwien.ac.at;
fhlawats@nt.tuwien.ac.at).\newline\hspace*{2mm} D. Eiwen is with
NuHAG, Faculty of Mathematics, University of Vienna, 1090 Vienna,
Austria (e-mail: daniel.eiwen@univie.ac.at).\newline\hspace*{2mm}
H. Rauhut is with the Hausdorff Center for Mathematics and the
Institute for Numerical Simulation, University of Bonn, 53115
Bonn, Germany (e-mail:
rauhut@hcm.uni-bonn.de).\newline\hspace*{2mm} Digital Object
Identifier 10.1109/JSTSP.2010.2042410
}\vspace*{1.3mm}}

\author{Georg Taub\"{o}ck, {\it Member, IEEE}, Franz Hlawatsch, {\it Senior Member, IEEE}, Daniel Eiwen, {\it Student Member, IEEE}, and Holger Rauhut\vspace*{-3mm}}

\maketitle

\begin{abstract}
We consider the application of {\em compressed sensing}\/ (CS) to
the estimation of doubly selective channels within
pulse-shaping multicarrier systems (which include OFDM systems as
a special case). By exploiting sparsity in the delay-Doppler domain,
CS-based channel estimation allows for an
increase in spectral efficiency through a reduction of the number
of pilot symbols. For combating leakage effects that limit the
delay-Doppler sparsity, we propose a sparsity-enhancing basis
expansion and a method for optimizing the basis with or without
prior statistical information about the channel. We also present
an alternative CS-based channel estimator for (potentially)
strongly time-frequency dispersive channels, which is capable of
estimating the ``off-diagonal'' channel coefficients
characterizing intersymbol and intercarrier interference
(ISI/ICI). For this estimator, we propose a basis construction
combining Fourier (exponential) and prolate spheroidal sequences.
Simulation results assess the performance gains achieved by
the proposed sparsity-enhancing processing techniques and by explicit estimation of ISI/ICI channel coefficients.
\end{abstract}


\section{{Introduction}}\label{sec:intro}

The recently introduced principle and methodology of {\em
compressed sensing}\/ (CS) allows the efficient reconstruction of
sparse signals from a very limited number of measurements
(samples) \cite{Can06,Don06}. CS has gained a fast-growing
interest in applied mathematics and signal processing \cite{rice}.
In this paper, we apply CS to the estimation of doubly selective (doubly dispersive, doubly spread) channels.
We consider pulse-shaping multicarrier (MC) systems, which include orthogonal
frequency-division multiplexing (OFDM) as a special case
\cite{kozmol98,Bingham90}. OFDM is part of, or proposed for,
numerous wireless standards like WLANs (IEEE 802.11a,g,n,
Hiperlan/2), fixed broadband wireless access (IEEE 802.16),
wireless personal area networks (IEEE 802.15),
digital audio and video broadcasting (DAB, DRM, DVB),
and future cellular communication systems (3GPP LTE) \cite{ieee802.11,ieee802.16,dvb01,dab95,etsi_drm,umts_lte}.

Coherent detection in such systems requires channel state
information (CSI) at the receiver. Usually, CSI is obtained by
embedding pilot symbols in the transmit signal and using a
least-squares (LS) \cite{jian01} or minimum mean-square error
(MMSE) \cite{li98} channel estimator. More advanced channel
estimators for MC transmissions include estimators employing
one-dimensional (1-D), double 1-D, or two-dimensional (2-D) MMSE
filtering algorithms \cite{Edfors98,Hoeher:1997aa,li-est00}; 2-D
irregular sampling techniques \cite{Fertl:2006aa}; or basis
expansion models \cite{leus_eusipco04,Borah99,zemen_sp05}. The
CS-based (``compressive'') channel estimation methodology proposed
in this paper exploits the fact that doubly selective multipath
channels tend to be dominated by a relatively small number of
clusters of significant paths, especially for large signaling
bandwidths and durations \cite{Raghavan07}. Conventional methods
for channel estimation do not take advantage of this
\emph{inherent sparsity} of the channel.  In
\cite{GT_icassp08,GT_eusipco08}, we proposed CS-based techniques
for estimating doubly selective channels within MC systems. We
demonstrated that CS provides a way to exploit channel sparsity in
the sense that the number of pilot symbols that have to be
transmitted for accurate channel estimation can be reduced.
Transmitting fewer pilots leaves more symbols for transmitting
data, which yields an increase in spectral efficiency.

For sparse channel estimation, several other authors have
independently proposed the application of CS methods or methods
inspired by the literature on sparse signal representations
\cite{Cotter02,Raghavan07,rabaste-channel,WLi07,Bajwa08a,sharp-scaglione_icassp08,Bajwa08b,Bajwa08c,Bajwa09}.
Both \cite{Cotter02} and \cite{WLi07} considered single-carrier
signaling and proposed variants of the matching pursuit algorithm
\cite{Mallat93} for channel estimation. The results were primarily
based on simulation and experimental implementations, without a CS
theoretical background. The channel estimation techniques
presented in \cite{Cotter02,Bajwa08a,sharp-scaglione_icassp08}
limited themselves to sparsity in the delay domain, i.e., they did
not exploit Doppler sparsity. The recent work in \cite{Bajwa08b}
and its extension to multiple-input/multiple-output (MIMO)
channels \cite{Bajwa08c}, on the other hand, considered both MC
signaling (besides single-carrier signaling) and sparsity in the
delay-Doppler domain, somewhat similar to \cite{GT_icassp08};
however, a different CS recovery technique was used. In
\cite{berger_09}, it is shown experimentally for MC communications over underwater acoustic channels
that compressive channel estimation outperforms traditional subspace
algorithms (root-MUSIC and ESPRIT).

In this paper, extending our work in
\cite{GT_icassp08,GT_eusipco08}, we present CS-based techniques
for estimating doubly selective channels that are potentially
strongly time- and/or frequency-dispersive. In MC systems, strong
channel dispersion may cause intersymbol interference (ISI) and/or
intercarrier interference (ICI) \cite{kozmol98}. One of the
proposed techniques enables the estimation of ISI/ICI channel
coefficients. We first present a basic compressive estimator for
mildly dispersive channels that yields estimates of the
``diagonal'' channel coefficients. Our focus is on \emph{leakage
effects} that limit the delay-Doppler sparsity, and which have not
been considered in
\cite{Cotter02,Raghavan07,rabaste-channel,WLi07,Bajwa08a,sharp-scaglione_icassp08,Bajwa08b,Bajwa08c,Bajwa09}.
For combating leakage effects and, hence, enhancing sparsity, we
then replace the discrete Fourier transform (DFT) used in
conventional compressive channel estimation by a more suitable
basis expansion. We also develop an iterative basis-optimization
procedure that is similar in spirit---but not algorithmically---to
dictionary learning techniques recently proposed in
\cite{ahelbr06,krmura03,Gribonval08}. This procedure is able to
take into account prior statistical information about the channel.
Next, we present an alternative compressive method for estimating
also the ``off-diagonal'' ISI/ICI channel coefficients of
potentially strongly dispersive channels (e.g., highly mobile
wireless channels or underwater acoustic channels
\cite{WLi07,berger_09}). Here, motivated by
\cite{Sle78,zemen_sp05}, we propose a sparsity-enhancing basis
expansion that combines Fourier (exponential) and prolate
spheroidal sequences.

This paper is organized as follows.
In Section \ref{sec:system_model}, we describe the MC system model.
In Section \ref{sec:cs_chest}, we present the basic compressive estimator for mildly dispersive channels.
An analysis of delay-Doppler leakage and its effect on delay-Doppler sparsity is performed in Section \ref{sec:sparsity}.
A sparsity-enhancing basis expansion and a framework and iterative algorithm for
optimizing the basis (with or without prior statistical information about the channel)
are developed in Sections \ref{sec:basis_expansion} and \ref{sec:basis_opt}, respectively.
In Section \ref{sec:disp_channel}, we propose a compressive
estimator and a basis expansion for (potentially) strongly dispersive channels.
Finally, simulation results presented in Section \ref{sec:simus} assess the performance gains achieved by
the proposed sparsity-enhancing basis expansions and by the estimation of ISI/ICI channel coefficients.

\section{Multicarrier System Model}\label{sec:system_model}

We assume a \emph{pulse-shaping} MC system for the sake of
generality and because of its advantages over conventional
cyclic-prefix (CP) OFDM
\cite{kozmol98,liu2004otf,schniter_CISS-06,Das_Schniter2007msi,Matz-Charlypaper07}.
This framework includes CP-OFDM as a special case. The complex
baseband domain is considered throughout.

\subsection{Modulator, Channel, Demodulator}\label{subsec:MC_mod}

The MC modulator generates the discrete-time transmit signal \cite{kozmol98}
\begin{equation}
\label{trans-symbol}
 s[n] \,=\ist \esum{l=0}{L-1}\esum{k=0}{K-1} a_{l,k}\,g_{l,k}[n] \,,
\end{equation}
where $L$ and $K$ denote the numbers of transmitted MC symbols and
subcarriers, respectively; $a_{l,k} \!\in\! \mathcal{A}$ ($l \rmv=\rmv 0,\dots,L\!-\!1$; $k \rmv=\rmv 0,\dots,K\!-\!1$)
denotes the complex data symbols, drawn from a finite symbol alphabet $\mathcal{A}$; and
$g_{l,k}[n] \triangleq g[n \rmv-\rmv lN] \, e^{\jmath 2\pi k(n-lN)/K}$ is a time-fre\-quen\-cy shift of a transmit pulse $g[n]$ ($N
\!\ge\! K$ is the symbol duration). Using an interpolation filter
with impulse response $f_1(t)$, $s[n]$ is converted into the continuous-time transmit
signal
\begin{equation}
\label{interp_f1}
s(t) \,=\rmv \sum_{n=-\infty}^{\infty} \! s[n] \ist f_1(t \rmv-\rmv  nT_{\rmv{\rm s}}) \,,
\end{equation}
where $T_{\rmv{\rm s}}$ is the sampling period. This signal is transmitted over a noisy, doubly selective channel,
at whose output the receive signal
\begin{equation}
\label{io_channel_impresp_cont}
    r(t) \eq \! \int_{-\infty}^\infty \! h(t,\tau) \ist s(t\rmv-\rmv \tau) \ist d\tau \ist+\ist z(t)
\end{equation}
is obtained. Here, $h(t,\tau)$ is the channel's time-varying impulse response
and $z(t)$ is complex noise. At the receiver, $r(t)$ is converted into the discrete-time receive signal
\begin{equation}
\label{rec-discr}
r[n] \,= \int_{-\infty}^\infty r(t) \ist f_2(n T_{\rmv{\rm s}} \rmv-\rmv t) \ist dt \,,
\end{equation}
where $f_2(t)$ is the impulse response of an anti-aliasing filter. Subsequently,
the MC demodulator calculates the ``demodulated symbols''
\begin{align}
  r_{l,k} &\eq  \langle r,\gamma_{l,k} \rangle \,=\! \isum{n}  \! r[n] \ist \gamma^*_{l,k}[n] \,, \nonumber\\
   & \hspace*{15mm} l = 0,\dots,L\!-\!1 \,, \;\;k = 0,\dots,K\!-\!1 \,.
\label{rec-symbol}
\end{align}
Here, $\gamma_{l,k}[n] \triangleq \gamma[n - lN] \ist e^{\jmath 2\pi k(n-lN)/K}$ with a receive pulse $\gamma[n]$. Finally, the demodulated symbols
$r_{l,k}$ are equalized and quantized according to the data symbol alphabet $\mathcal{A}$.

Combining \eqref{interp_f1}--\eqref{rec-discr},
we obtain an equivalent discrete-time channel
that is described by the following relation between the discrete-time signals $s[n]$ and $r[n]$:
\be
\label{io_channel_impresp}
r[n] \eq \!\rmv \sum_{m=-\infty}^{\infty}\!\! h[n,m] \ist s[n\rmv-\rmv m] \ist+\ist z[n] \,,
\ee
with the discrete-time time-varying impulse response
$h[n,m] = \int_{-\infty}^\infty \int_{-\infty}^\infty h(t+nT_{\rmv{\rm s}}, \tau) \ist f_1(t \rmv-\rmv \tau+m T_{\rmv{\rm s}}) \ist f_2(-t) \, d t \ist d\tau$
and the discrete-time noise $z[n] = \int_{-\infty}^\infty z(t) \ist f_2(n T_{\rmv{\rm s}} \rmv-\rmv t) \ist dt$.

CP-OFDM is a simple special case of the pulse-shaping MC framework; it is obtained for a rectangular
transmit pulse $g[n]$ that is $1$ for $n = 0 , \dots, N\!-\!1$ and $0$
otherwise, and a rectangular receive pulse $\gamma[n]$ that is $1$
for $n = N\!-\!K, \dots,N\!-\!1$ and $0$ otherwise ($N\!-\!K \geq 0$ is the CP length).

\subsection{System Channel}\label{sec:io}

Next, we consider the equivalent \emph{system channel} that subsumes the MC modulator, interpolation filter, physical channel,
anti-aliasing filter, and MC demodulator.
Combining \eqref{rec-symbol}, \eqref{io_channel_impresp}, and \eqref{trans-symbol}, we obtain
\begin{align}
  r_{l,k} &\eq \sum_{l'=0}^{L-1} \sum_{k'=0}^{K-1} H_{l,k;l'\!,k'}\, a_{l'\!,k'} \ist+\ist z_{l,k} \,, \nonumber\\
   & \hspace*{15mm} l = 0,\dots,L\!-\!1 \,, \;\;k = 0,\dots,K\!-\!1 \,,
\label{gen.io.rel_4D}
\end{align}
with $z_{l,k} = \langle z, \gamma_{l,k} \rangle = \isum{n}  z[n] \ist \gamma^*_{l,k}[n]$. The system channel coefficients $H_{l,k;l'\!,k'}$
describe ICI for $k \rmv\not=\rmv k'\rmv$ and $l \!=\! l'\rmv$ and ISI for $l \rmv\not=\rmv l'$; they can be expressed in terms of
$h[n,m]$, $g[n]$, and $\gamma[n]$ \cite{kozmol98}.

Let $\gamma[n]$ be zero outside $\{0,\ldots,L_\gamma\}$. To compute $r_{l,k}$ in (\ref{rec-symbol}) for $l = 0,\dots,L\!-\!1$, we need to know $r[n]$ for
$n=0,\ldots, N_{r}\!-\!1$, where $N_{r} \triangleq (L\!-\!1)N + L_\gamma + 1$.
In this interval, we can rewrite \eqref{io_channel_impresp} as
\begin{equation}
\label{io_channel_spreading}
    r[n] \,=\!  \sum_{m=-\infty}^{\infty} \sum_{i=0}^{N_{r}-1} \!S_h[m,i] \ist s[n\rmv-\rmv m] \ist e^{\jmath 2 \pi \frac{ni}{N_{r}}} +\,z[n] \,,
\end{equation}
with the {\em discrete-delay-Doppler spreading function}\/
\cite{Bel63}
\be
\label{Spread_fct}
S_h[m,i] \,\triangleq\, \frac{1}{N_{r}} \! \sum_{n=0}^{N_{r}-1} \! h[n,m] \ist e^{-\jmath 2 \pi\frac{in}{N_{r}}} \,, \quad\; m, i \in \mathbb{Z} \,,
\end{equation}
which represents the channel in terms of discrete delay (time
shift) $m$ and discrete Doppler frequency shift $i$. Combining
\eqref{rec-symbol}, \eqref{io_channel_spreading}, and
\eqref{trans-symbol}, and assuming that $h[n,m]$ is causal with
maximum delay at most $K\!-\!1$, i.e., $h[n,m] = 0$ for $m \not\in
\{ 0,\dots,K\!-\!1\}$, we reobtain the system channel relation
\eqref{gen.io.rel_4D}, however with the system channel
coefficients $H_{l,k;l'\!,k'}$ now expressed in terms of the
delay-Dopler representation $S_h[m,i]$. Specializing this
expression to $(l'\!,k') \!=\! (l,k)$ and using the approximation
$N_{r} \rmv\approx\rmv LN$ (which is exact for CP-OFDM) yields the
following expression for the \emph{diagonal} channel coefficients
$H_{l,k} \triangleq H_{l,k;l,k}\ist$ ($L$ is assumed even for
mathematical convenience):
\begin{align}
  H_{l,k} &\,=\ist \sum_{m=0}^{K-1} \sum_{i=-L/2}^{L/2-1} \!F[m,i] \, e^{-\jmath 2 \pi (\frac{km}{K}-\frac{li}{L} )} \ist, \nonumber\\
   & \hspace*{15mm} l = 0,\dots,L\!-\!1 \,, \;\;k = 0,\dots,K\!-\!1 \,,
\label{System_Channel_Det_1}
\end{align}
with
\be
F[m,i] \,\triangleq\ist \sum_{q=0}^{N-1} \rmv S_h[m,i+qL] \, A^{*}_{\gamma,g}\rmv \Big( m, \frac{i+qL}{N_{r}} \Big) \,, \quad\; i \in \mathbb{Z} \,.
\label{F_S_A}
\ee
Here, $A_{\gamma,g}(m,\xi) \triangleq \isum {n} \gamma[n] \, g^*[n\rmv-\rmv m] \ist e^{-\jmath 2\pi \xi n}$
is the {\em cross-ambiguity function}\/ \cite{fla-book2} of $\gamma[n]$ and $g[n]$.

\section{Compressive Channel Estimation}\label{sec:cs_chest}

We now present the basic compressive channel estimation method \cite{GT_icassp08,Bajwa08b}. This method
enables estimation of the diagonal channel coefficients $H_{l,k} = H_{l,k;l,k}\ist$, which is sufficient for
mildly dispersive channels.

\subsection{Pilot-assisted Channel Estimation}\label{subsec:PB-ChannelEst}

Our goal is to estimate the system channel coefficients $H_{l,k} = H_{l,k;l,k}$ from the system channel output $r_{l,k}$, aided by some known pilot symbols.
For practical (underspread \cite{Bel63}) wireless channels and practical transmit and receive
pulses, $F[m,i]$ in \eqref{F_S_A} is effectively supported in a subregion of the delay-Doppler plane. Thus,
hereafter we assume that the support of $F[m,i]$ (within the
fundamental $i$ period $\{-L/2,\ldots,L/2\rmv-\!1\}$; note that $F[m,i]$ is $L$-periodic in $i$) is contained
in $\{ 0,\ldots,  D\!-\! 1\} \times \{-J/2,\ldots,J/2\rmv-\!1\}$, where $D\leq K$
and $J \le L$. Here, $J$ is chosen even, and $D$ and $J$ are such that $\Delta K \triangleq
K/D$ and $\Delta L \triangleq L/J$ are integers. Note that we also allow the limiting case of full support in either or both dimensions, that is,
$D=K$ (i.e., $\Delta K \rmv=\rmv 1$) and/or $J=L$ (i.e., $\Delta L \rmv=\rmv 1$).
Because of \eqref{System_Channel_Det_1}, the $H_{l,k}$ are then uniquely specified by their values on the \emph{subsampled time-frequency grid}
\begin{align*}
  \mathcal{G} &\,\triangleq\, \{(l,k) = (\lambda\,\Delta L, \kappa\,\Delta K):\,\,\lambda\rmv=0,\ldots,J\!-\!1\,,\\
   & \hspace*{50mm} \kappa\rmv\rmv=0,\ldots, D\!-\!1\} \,.
\end{align*}
These subsampled values are given by
\begin{align}
  H_{\lambda\,\Delta L, \kappa\,\Delta K} &\eq \rmv\sum_{m=0}^{D-1}\sum_{i=-J/2}^{J/2-1} \! F[m,i] \,
    e^{-\jmath 2 \pi ( \frac{\kappa m}{D}-\frac{\lambda i}{J} )} \ist , \nonumber\\
   & \hspace*{5mm}  \lambda=0,\ldots,J\!-\!1\,, \;\; \kappa=0,\ldots, D\!-\!1 \,.
\label{System_Channel_Det_subs}
\end{align}
The time-frequency subsampling is desirable because it reduces the dimensionality of the estimation problem, and thus
tends to result in better estimation performance.

Suppose now that pilot symbols $a_{l,k} \!=\rmv p_{l,k}$ are transmitted at
time-frequency positions $(l,k) \!\in\! \mathcal{P}\rmv$, where $\mathcal{P} \!\subset\! \mathcal{G}$, i.e.,
the \emph{pilot position set} $\mathcal{P}$ is a subset of the subsampled time-frequency grid $\mathcal{G}$.
For mildly dispersive channels, the ISI and ICI are small.
Then, at the pilot positions $(l,k) \!\in\! \mathcal{P}$, it is convenient to rewrite the system channel relation \eqref{gen.io.rel_4D} as
$r_{l,k} =H_{l,k}\, p_{l,k} + \tilde z_{l,k}$, where all ISI and ICI are now subsumed by the noise/interference
term $\tilde z_{l,k}$. Based on this relation and the known $p_{l,k}$, the receiver
calculates channel coefficient estimates $\hat{H}_{l,k}$ at the pilot positions according to
\be
    \hat{H}_{l,k} \,\triangleq\, \frac{r_{l,k}}{p_{l,k}} \eq H_{l,k} \,+\, \frac{\tilde z_{l,k}}{p_{l,k}} \;, \qquad (l,k) \in \mathcal{P} \,.
\label{pilot_ch_est} \ee The last expression shows that the
$H_{l,k}$ for $(l,k) \in \mathcal{P}$ are known up to additive
noise/interference terms $\tilde z_{l,k}/p_{l,k}$. A conventional
channel estimator then uses some interpolation technique to
calculate channel estimates $\hat{H}_{l,k}$ for all $(l,k)$ from
the $\hat{H}_{l,k}$ for $(l,k) \in \mathcal{P}$ (e.g.,
\cite{jian01,li98,Edfors98,Hoeher:1997aa,li-est00,Fertl:2006aa}).
In contrast, the proposed compressive channel estimator uses a CS
recovery technique to obtain an estimate of $F[m,i]$ and, in turn,
of the $H_{l,k}$.

\subsection{Some CS Fundamentals}\label{sec:comp_sens}

Before presenting the CS-based channel estimator, we need to review some CS fundamentals \cite{Can06,Don06}.
CS considers the \emph{sparse reconstruction problem} of
estimating an (approximately) sparse vector $\mathbf{x} \in
\C^{\M}$ from an observed vector of measurements $\mathbf{y}
\!\in\! \C^{Q}$ based on the linear model (``measurement
\vspace{-1mm} equation'')
\be
    \mathbf{y} \ist =\ist \mathbf{\Phi} \mathbf{x} + \mathbf{z} \,.
\label{sparse_reconstr}
\ee
Here, $\mathbf{\Phi} \!\in\! \C^{Q \times\M}$ is a known measurement matrix and $\mathbf{z} \!\in\! \C^{Q}$ is an unknown vector
that accounts for measurement noise and modeling errors. The reconstruction is subject to the constraint that $\mathbf{x}$
is (approximately) $S$-\emph{sparse}, i.e., at most $S$ of its
entries are not (approximately) zero. The positions (indices) of the
significantly nonzero entries of $\mathbf{x}$ are unknown.
Typically, the number of variables to be estimated is much larger
than the number of measurements, i.e., $\M \!\gg\rmv Q$.
Thus, $\mathbf{\Phi}$ is a fat matrix.

We briefly review some CS recovery methods.
\emph{Basis pursuit} (BP) \cite{chdosa99,Can06a} and {\em orthogonal matching pursuit}\/ (OMP) \cite{Tropp04} are probably the most popular ones.
Whereas for BP theoretical performance guarantees are available, OMP lacks similar results. However,
OMP allows a faster implementation, and simulation results even demonstrate a better performance. Low computational
complexity is important since the channel has to be estimated in real time. \emph{CoSaMP} \cite{netrXX} allows an even
faster implementation than OMP. (Note that \emph{subspace pursuit}
\cite{dami08} is a very similar method.) Using an efficient
implementation of the pseudoinverse by means of the LSQR algorithm
 \cite{paigesaunders82}, we observed a run time that was only less than half that of OMP, and a performance that was only slightly poorer.
An advantage of CoSaMP is the availability of performance bounds. Hence, CoSaMP offers a good compromise between low
complexity, good practical performance, and provable performance guarantees.

The performance guarantees of BP and CoSaMP are phrased as an
upper bound on the approximation error ${ \| \hat{\mathbf{x}}
\rmv-\rmv \mathbf{x} \|}_2$, where $\hat{\mathbf{x}}$ denotes the
estimate of $\mathbf{x}$. This bound is valid if the measurement
matrix $\mathbf{\Phi}$ satisfies $(1\!-\!\delta) \, {\| \mathbf{x}
\|}_{2}^{2} \leq \left\| \mathbf{\Phi} \mathbf{x} \right
\|_{2}^{2} \leq (1\!+\rmv\delta) \, {\| \mathbf{x} \|}_{2}^{2}$
for all $S$-sparse vectors $\mathbf{x} \rmv\in \C^M\!$, with some
positive constant $\delta$. This is known as the \emph{restricted
isometry property} (RIP), and the smallest $\delta$ is termed the
\emph{restricted isometry constant} $\delta_{S}$. For a small
bound on ${ \| \hat{\mathbf{x}} \rmv-\rmv \mathbf{x} \|}_2$,
$\delta_{S}$ should be small. It has been shown
\cite{Can06,rud06,ra08-1} that if $\mathbf{\Phi} \!\in\! \C^{Q
\times \M}\rmv$ is constructed by selecting uniformly at random
$Q$ rows\footnote{That is, all possible choices of $Q$ rows are
equally likely.} from a unitary $\M \!\times\rmv \M\rmv$ matrix
$\mathbf{U}$ and normalizing the columns (so that they have unit
$\ell_2$ norms), a sufficient condition for $\mathbf{\Phi}$ to
satisfy the RIP with a restricted isometry constant that is
bounded as $\delta_{S}\leq \gamma$ with probability $1\!-\rmv
\eta$ is provided by the following lower bound on the number of
observations:
\begin{equation}
\label{UUP_satisfy_suffcond}
    Q \,\geq\, C \, \gamma^{-2} \ist (\ln \M)^{4} \ist \mu_\mathbf{U}^{2} \ist S \ist \ln(1/\eta) \,.
\end{equation}
Here, $\mu_\mathbf{U} \triangleq \sqrt{\M }\max_{i, j}|U_{i,j}|$ (known as
the {\em coherence}\/ of $\mathbf{U}$) and $C $ is a constant.

Further CS recovery methods include {\em thresholding}
\cite{Kunis08}, the {\em stagewise OMP} \cite{Donoho06}, the {\em
LARS method} \cite{efhajoti04,Donoho06a}, the {\em Lasso}
\cite{tibshirani_96,Loris09} (equivalent to \emph{BP denoising}
\cite{Loris09}), and Bayesian methods
\cite{ji-xue-carin_TSP-08,schniter_BayMatchPur08}. In
\cite{Bajwa08b,Bajwa08c}, the {\em Dantzig selector} (DS)
\cite{Candes07} was applied to sparse channel estimation. DS
satisfies optimal asymptotic performance bounds when the noise
vector $\mathbf{z}$ is modeled as random. However, for the
practically relevant case of finite (moderate) $Q$ and $M$, the
performance of DS is not necessarily superior. In our experiments,
we did not observe any performance or complexity advantages of DS
over BP, OMP, and CoSaMP.

\subsection{Basic Compressive Channel Estimator}\label{sec:chest}

We now combine pilot-assisted channel estimation with CS recovery. The central assumption of compressive channel estimation
is that $S_h[m,i]$ is ``compressible'' \cite{Can06a} or approximately $S$-sparse, i.e., at most $S$ values of $S_h[m,i]$ (in the fundamental
$i$ period $\{-L/2,\ldots,L/2\rmv-\!1\}$) are not approximately zero. This
approximate ``delay-Doppler sparsity'' assumption will be further considered in Section \ref{sec:sparsity}. Note that it implies that also
$F[m,i] = \sum_{q=0}^{N-1} S_h[m,i+qL] \, A^{*}_{\gamma,g}\big( m, \frac{i+qL}{N_{r}} \big)$ is approximately $S$-sparse.

Our starting-point is the 2-D DFT relation
(\ref{System_Channel_Det_subs}), which can be written as the 2-D expansion
\be
\label{BEM_dft}
  H_{\lambda\,\Delta L,\kappa  \Delta K} \eq\rmv \sum_{m=0}^{D-1}\sum_{i=-J/2}^{J/2-1} \!\alpha_{m,i} \, u_{m,i}[\lambda,\kappa ] \,,
\ee
with $\alpha_{m,i} \triangleq \sqrt{JD} \, F[m,i]$ and $u_{m,i}[\lambda,\kappa ] \triangleq (1/\sqrt{JD}) \, e^{-\jmath 2 \pi (\kappa m/D- \lambda i/J )}$.
The functions $H_{\lambda\,\Delta L,\kappa\,\Delta K}$ and $u_{m,i}[\lambda,\kappa]$
are defined for $\lambda=0,\ldots,J\!-\!1$ and $ \kappa =0,\ldots, D\!-\!1$ and may thus be considered as
$J \times D$ matrices. Define the vectors $\mathbf{h} \triangleq \mbox{vec} \big\{ H_{\lambda\,\Delta L,\kappa\,\Delta K}\big\}$ and
$\mathbf{u}_{m,i} \triangleq \mbox{vec} \big\{ u_{m,i}[\lambda,\kappa] \big\}$ of length $JD$
by stacking all columns of these matrices (e.g., $\mathbf{h} = [h_{1} \cdots\ist h_{JD}]^{T}$ with $h_{\kappa J + \lambda +1} = H_{\lambda\,\Delta L,\,\kappa\,\Delta K}$).
We can then rewrite (\ref{BEM_dft}) as
\be
\label{ch_est}
    \mathbf{h} \eq\rmv \sum_{m=0}^{D-1}\sum_{i=-J/2}^{J/2-1} \!\alpha_{m,i} \ist \mathbf{u}_{m,i}
    \eq \mathbf{U} \aalpha\,,
\ee
where $\aalpha \triangleq \mbox{vec} \big\{ \alpha_{m,i} \big\}$ and $\mathbf{U}$ is the $JD \times JD$ matrix
whose $\big((i + J/2) D + m + \!1\big)\ist$th column is given by the vector $\mathbf{u}_{m,i}$.
Because the $\mathbf{u}_{m,i}$ are orthonormal, $\mathbf{U}$ is a unitary matrix.

According to Section \ref{subsec:PB-ChannelEst}, there are $|\mathcal{P}|$ pilot symbols at time-frequency positions $(l,k)
\!\in\! \mathcal{P}$. Thus, $|\mathcal{P}|$ of the $JD$ entries of
$\mathbf{h}$ are given by the channel coefficients $H_{l,k}$ at the pilot positions $(l,k) \!\in\! \mathcal{P}$. Let $\mathbf{h}^{({\rm p})}\!$ denote the
corresponding length-$|\mathcal{P}|$ subvector of $\mathbf{h}$,
and let $\mathbf{U}^{({\rm p})}$ denote the $|\mathcal{P}|
\rmv\times\! JD$ submatrix of $\mathbf{U}$ constituted by the
corresponding $|\mathcal{P}|$ rows of $\mathbf{U}$. Reducing
\eqref{ch_est} to the pilot positions, we obtain
\be
\label{ch_est_pilot_1}
   \mathbf{h}^{({\rm p})} \rmv\eq \mathbf{U}^{({\rm p})} \aalpha \eq \mathbf{\Phi} \mathbf{x} \,,
\ee
with $\mathbf{\Phi} \triangleq\rmv \sqrt{ \frac{JD}{|\mathcal{P}|} } \,\mathbf{U}^{({\rm p})}$ and $\mathbf{x} \triangleq\rmv \sqrt{ \frac{|\mathcal{P}|}{JD} } \, \aalpha$.
Note that $\mathbf{\Phi}$ is normalized such that its columns
have unit $\ell_2$-norm, and that the length-$JD$ vector
$\mathbf{x}$ is, up to a constant factor, the vector form of $F[m,i]$.

Our task is to estimate $\mathbf{x}$ based on relation \eqref{ch_est_pilot_1}. The vector $\mathbf{h}^{({\rm p})}$ is unknown,
but we can approximate it by the corresponding vector of pilot-based channel coefficient estimates $\hat{H}_{l,k} \big|_{(l,k) \in \mathcal{P}}\ist$
(see \eqref{pilot_ch_est}).
For consistency with the notation used in Section \ref{sec:comp_sens}, this latter vector will be denoted as $\mathbf{y}$ (rather than $\hat{\mathbf{h}}^{({\rm p})}$).
According to \eqref{pilot_ch_est}, $\mathbf{y} = \mathbf{h}^{({\rm p})} \rmv + \mathbf{z}$,
where $\mathbf{z}$ is the vector of noise/interference terms $\tilde z_{l,k}/p_{l,k} \big|_{(l,k) \in \mathcal{P}}\ist$. Inserting \eqref{ch_est_pilot_1},
we finally obtain the \emph{measurement equation}
\be
\mathbf{y} \eq \mathbf{\Phi} \mathbf{x} \ist+\ist \mathbf{z} \,.
\label{chest_sparse_reconstr}
\ee
The vector $\mathbf{x}$ is approximately $S$-sparse because $S_h[m,i]$ was assumed approximately $S$-sparse.
Thus, \eqref{chest_sparse_reconstr} is seen to be a sparse reconstruction problem of the form \eqref{sparse_reconstr}, with
dimensions $M \rmv=\rmv \text{dim} \{ \mathbf{x} \} \rmv=\rmv JD$ and $Q  \rmv=\rmv \text{dim} \{ \mathbf{y} \} \rmv=\rmv |\mathcal{P}|$ and sparsity $S$.
We can hence use one of the CS recovery techniques reviewed in Section \ref{sec:comp_sens} to obtain an estimate of
$\mathbf{x}$ or, equivalently, of $\aalpha = \sqrt{\frac{JD}{|\mathcal{P}|}} \, \mathbf{x}$ or of
$F[m,i] = \frac{\alpha_{m,i}}{\sqrt{JD}}$. From the estimate $\hat F[m,i]$ of $F[m,i]$,
estimates of all channel coefficients $H_{l,k}$ are finally obtained via (\ref{System_Channel_Det_1}).

According to its definition $\mathbf{\Phi} =\rmv \sqrt{ \frac{JD}{|\mathcal{P}|} } \,\mathbf{U}^{({\rm p})}$,
the measurement matrix $\mathbf{\Phi}$ is constructed by selecting $|\mathcal{P}|$ rows
of the unitary $JD \!\times\rmv JD$ matrix $\mathbf{U}$ and
normalizing the resulting columns. This agrees with the
construction of $\mathbf{\Phi}$ described in Section \ref{sec:comp_sens} in the context of BP and CoSaMP. To
be fully consistent with that construction, we have to select the
$|\mathcal{P}|$ rows of $\mathbf{U}$ uniformly at random.
The indices of these rows equal the $|\mathcal{P}|$ indices within the index range $\{1,\ldots,JD \}$
of the channel vector $\mathbf{h}$ that correspond to the set of pilot positions $\mathcal{P}$. We conclude that the pilot positions $(l,k) \!\in\! \mathcal{P}$ have to be
selected uniformly at random within the subsampled time-frequency grid $\mathcal{G}$, in the sense that the $|\mathcal{P}|$ ``pilot indices''
within the index range $\{1,\ldots,JD \}$ of $\mathbf{h}$ are selected uniformly at random.

For BP and CoSaMP, in order to achieve a small upper bound on the
reconstruction error ${ \| \hat{\mathbf{x}} \rmv-\rmv \mathbf{x}
\|}_2$ as discussed in Section \ref{sec:comp_sens}, the number of
pilots should satisfy condition (\ref{UUP_satisfy_suffcond}). In
our case, this (sufficient) condition becomes
\[
    |\mathcal{P}| \,\geq\, C \, \gamma^{-2} \ist \big( \rmv\ln (JD) \big)^{4} \, S \ist \ln(1/\eta) \,,
\]
with an appropriately chosen $\gamma$ (note that $\mu_\mathbf{U}
\!=\! 1$). This bound suggests that the required number of pilots
scales at most linearly with the delay-Doppler sparsity parameter
$S$ and poly-logarithmically with the system design parameters $J$
and $D$. Note that the pilot positions are randomly chosen (and
communicated to the receiver) before the beginning of data
transmission; they are fixed during data transmission.

\section{Delay-Doppler Sparsity and Leakage Effect}\label{sec:sparsity}

In this section, we analyze the sparsity of the channel's
delay-Doppler representation for a simple time-varying multipath
channel model comprising $P$ specular (point) scatterers with
fixed delays $\tau_p$ and Doppler frequency shifts $\nu_p$ for $p
= 1,\dots,P$. This simple model is often a good approximation to
real mobile radio channels \cite{BaSca-book00,GM_eusipco_06}. The
channel impulse response thus has the form \be
\label{dominant_impresp_cont} h(t,\tau) \eq\rmv \sum_{p=1}^{P}
\eta_{p} \, \delta(\tau\!-\!\tau_{p}) \, e^{\jmath 2\pi \nu_{p} t
} , \ee where $\eta_{p}$ characterizes the attenuation and initial
phase of the $p$th propagation path and $\delta(\cdot)$ is the
Dirac delta. The discrete-delay-Doppler spreading function
(\ref{Spread_fct}) then becomes
\begin{align}
S_h[m,i] &\eq \frac{1}{N_{r}} \sum_{p=1}^{P} \eta_{p} \, \phi^{(\nu_{p})}\Big(m  - \frac{\tau_{p}}{T_{\rmv{\rm s}}}\Big)
  \sum_{n=0}^{N_{r}-1} e^{\jmath 2\pi (\nu_{p} T_{\rmv{\rm s}} - \frac{i}{N_{r}})n } \nonumber\\
&\eq \sum_{p=1}^{P} \rmv\eta_{p} \, e^{\jmath \pi (\nu_{p} T_{\rmv{\rm s}} - \frac{i}{N_{r}}) (N_{r}-1) }  \nonumber\\[-2mm]
& \hspace{20mm} \times \Lambda^{(\nu_{p})}\Big( m -\rmv \frac{\tau_p}{T_{\rmv{\rm s}}}, \ist i -\rmv \nu_{p} T_{\rmv{\rm s}} N_{r} \rmv \Big) \,,
\label{Spread_fct_dom}
\end{align}
with
\[
\Lambda^{(\nu)}(x,y) \,\triangleq\, \phi^{(\nu)}(  x ) \ist \psi(  y ) \,,
\]
where
\begin{align}
\phi^{(\nu)}(x) &\,\triangleq\, \int_{-\infty}^{\infty} \! f_1( T_{\rmv{\rm s}} \ist x \rmv-\rmv t) \ist f_2(t) \, e^{-\jmath 2\pi \nu t}\ist dt
  \label{phi_def}\\[0mm]
\psi(y) &\,\triangleq\, \frac{1}{N_{r}} \, e^{\jmath \pi \frac{y}{N_{r}} (N_{r}-1) } \sum_{n=0}^{N_{r}-1} e^{-\jmath 2 \pi \frac{y}{N_{r}} n}  \nonumber\\[0mm]
& \eq \frac{\sin(\pi y)}{N_{r}\sin(\pi y/N_{r})} \;.
\label{psi_def}
\end{align}
It is seen from \eqref{Spread_fct_dom} that, although we assumed
specular scattering, $S_h[m,i]$ does not consist of Dirac-like
functions at the delay-Doppler points of the scatterers,
$(\tau_p/T_{\rmv{\rm s}}, \nu_{p} T_{\rmv{\rm s}} N_{r})$. Rather,
there occurs a \emph{leakage effect} which is characterized by the
function $\Lambda^{(\nu)}(x,y) = \phi^{(\nu)}(  x ) \ist \psi( y
)$, and which is stronger for a broader $\Lambda^{(\nu)}(x,y)$.
The leakage effect is due to the finite transmit bandwidth
($\approx 1/T_{\rmv{\rm s}}$) and the finite blocklength ($N_r
\approx LN$). It is important for compressive channel estimation
because it implies a poorer sparsity of $S_h[m,i]$. Note that
whereas a large blocklength reduces the leakage effect, it also
implies that the specular model with constant parameters
\eqref{dominant_impresp_cont} is a less accurate approximation
and, thus, that the continuous-delay-Doppler spreading function
\cite{Bel63} is less sparse. This motivates an extension of the
compressive channel estimation method that is able to reduce the
leakage effect (see Section \ref{sec:basis_expansion}).

In view of (\ref{Spread_fct_dom}), studying the sparsity of
$S_h[m,i]$ essentially amounts to studying the sparsity of
$\Lambda^{(\nu_{p})}( m -\rmv \tau_p/T_{\rmv{\rm s}}, \ist i -\rmv
\nu_{p} T_{\rmv{\rm s}} N_{r} ) = \phi^{(\nu_{p})}( m -\rmv
\tau_p/T_{\rmv{\rm s}} ) \, \psi(  i -\rmv \nu_{p} T_{\rmv{\rm s}}
N_{r} )$. To this end, we first consider the energy of those
samples of $\phi^{(\nu_{p})}(m \rmv-\rmv \tau_p/T_{\rmv{\rm s}}
)$ whose distance from $\tau_p/T_{\rmv{\rm s}}$ is greater than
$\Delta m \in \{1,2,\dots\}$, i.e., $|m-\tau_{p}/T_{\rmv{\rm s}}|
> \Delta m$. We assume that $\phi^{(\nu)}(x)$ exhibits at least a
polynomial decay, i.e., $|\phi^{(\nu)}(x)|\leq C \ist
(1+|x/x_{0}|)^{-s}$ with $s\rmv\ge\! 1$, for some positive
constants $C$ and $x_{0}$. This includes the following important
special cases: (i) the ideal lowpass filter, i.e., $f_1(t) =
f_2(t) =\sqrt{1/T_{\rmv{\rm s}}}\ist \sinc( t / T_{\rmv{\rm s}})$
with $\sinc(x) \triangleq \frac{\sin (\pi x)}{\pi x}$, here $s
\rmv=\rmv 1$; and (ii) the family of root-raised-cosine filters:
if both $f_1(t)$ and $f_2(t)$ are equal to the root-raised-cosine
filter with roll-off factor $\rho$, then, for $\nu$ not too large,
$\phi^{(\nu)}(x) \approx \sinc(x) \ist \cos(\rho \pi x)/[1\!-\! (2
\rho x)^{2}]$ and $s=3$. Based on the polynomial-decay assumption,
one can show the following bound \cite{GT_eusipco08} on the energy
of all $\phi^{(\nu_{p})}(m \rmv-\rmv \tau_p/T_{\rmv{\rm s}}  )$
with $|m-\tau_{p}/T_{\rmv{\rm s}}| > \Delta m$:
\[
    \sum_{ |m-\tau_{p}/T_{\rmv{\rm s}}| > \Delta m } \! \Big|\phi^{(\nu_{p})} \Big(m - \frac{\tau_p}{T_{\rmv{\rm s}}} \Big)  \Big|^{2}
\ist\leq\, \frac{2 \ist C^{2} x_{0} }{2s -\! 1} \Big( 1 + \frac{\Delta m -\! 1}{x_{0}} \Big)^{\!\rmv -2s+1} .
\]
Hence, the energy of $\phi^{(\nu_{p})}(m \rmv-\rmv \tau_p/T_{\rmv{\rm s}}  )$ outside the interval $\big[\lfloor \tau_p/T_{\rmv{\rm s}} - \Delta m
\rfloor,\lceil \tau_p/T_{\rmv{\rm s}} + \Delta m \rceil\big]$ decays polynomially of order $2s-\!1$ with respect to $\Delta m$.

In a similar manner, we consider the energy of those samples of
$\psi(i - \nu_{p} T_{\rmv{\rm s}} N_{r} ) $ whose distance (up to
the modulo-$N_{r}$ operation, see below) from $\nu_{p} T_{\rmv{\rm
s}} N_{r}$ is greater than $\Delta i \in \{2,\ldots,\lfloor
N_{r}/2\rfloor\}$. Let $\mathcal{I}$ denote the set
$\{0,\ldots,N_{r}\!-\!1\}$ with the exception of all $i \rmv=\rmv
i_{\Z} \,\text{mod}\, N_{r}$, where $i_{\Z}$ is any integer with
$|i_{\Z}-\nu_{p} T_{\rmv{\rm s}} N_{r}|\leq \Delta i$. From
\eqref{psi_def}, one can obtain the bound \cite{GT_icassp08}
\[
\sum_{i \in \mathcal{I}} \big|\psi(i - \nu_{p} T_{\rmv{\rm s}} N_{r} )  \big|^{2}
\,\leq \, \frac{1}{\pi (\Delta i \rmv-\! 1)} \,,
\]
which shows that the energy of $\psi( i -\rmv \nu_{p} T_{\rmv{\rm s}} N_{r} )$ outside the interval $\big[\lfloor
\nu_{p} T_{\rmv{\rm s}} N_{r} - \Delta i \rfloor,\lceil \nu_{p}
T_{\rmv{\rm s}} N_{r} + \Delta i\rceil\big]$ (modulo $N_{r}$) decays linearly (polynomially of order $1$) with respect to $\Delta i$.

From these decay results, it follows that $\Lambda^{(\nu_{p})}( m -\rmv \tau_p/T_{\rmv{\rm s}}, \ist i -\rmv \nu_{p} T_{\rmv{\rm s}} N_{r})
= \phi^{(\nu_{p})}( m -\rmv \tau_p/T_{\rmv{\rm s}} ) \, \psi(  i -\rmv \nu_{p} T_{\rmv{\rm s}} N_{r} )$
can be considered as an \emph{approximately sparse} (or {\em compressible}, in CS terminology \cite{Can06a}) function.
Thus, as an approximation, we can model $\Lambda^{(\nu_{p})}( m -\rmv \tau_p/T_{\rmv{\rm s}}, \ist i -\rmv \nu_{p} T_{\rmv{\rm s}} N_{r} )$
as $N_{\hspace{-.15mm}\Lambda}$-sparse,
with an appropriately chosen sparsity parameter
$N_{\hspace{-.15mm}\Lambda}$. It then follows from
(\ref{Spread_fct_dom}) that $S_h[m,i]$ is $P N_{\hspace{-.15mm}\Lambda}$-sparse,
and the same is true for $F[m,i]$ in \eqref{F_S_A}.
Unfortunately, $N_{\hspace{-.15mm}\Lambda}$ cannot be chosen
extremely small because of the strong leakage that is due to the slowly (only linearly) decaying
factor $\psi( i -\rmv \nu_{p} T_{\rmv{\rm s}} N_{r} )$. This limitation motivates the introduction of a sparsity-enhancing basis expansion in the next section.

\section{Sparsity-Enhancing Basis Expansion}\label{sec:basis_expansion}

The 2-D DFT relation \eqref{System_Channel_Det_subs} underlying the basic compressive channel estimator
is an expansion of the subsampled channel coefficients $H_{\lambda\,\Delta L,\kappa\,\Delta K}$ into the
2-D DFT basis $u_{m,i}[\lambda,\kappa] = (1/\sqrt{JD}) \, e^{-\jmath 2 \pi ( \kappa m/D - \lambda i/J )}$ (see
\eqref{BEM_dft}). The sparsity of the expansion coefficients $\alpha_{m,i} = \sqrt{JD} \, F[m,i]$ was shown above to be limited by
the slowly (only linearly) decaying function $\psi( i -\rmv \nu_{p} T_{\rmv{\rm s}} N_{r} )$.
In order to enhance the sparsity, we now introduce a generalized 2-D expansion of $H_{\lambda\Delta L,\kappa  \Delta K}$ into orthonormal basis functions $v_{m,i}[\lambda,\kappa]$:
\begin{align}
  H_{\lambda\,\Delta L,\kappa\,\Delta K} &\eq\rmv \sum_{m=0}^{D-1}\sum_{i=-J/2}^{J/2-1} \!\beta_{m,i} \, v_{m,i}[\lambda,\kappa ] \,, \nonumber\\
   & \hspace*{5mm} \lambda =0,\ldots,J\!-\!1\,, \;\;  \kappa=0,\ldots, D\!-\!1 \,.
\label{BEM}
\end{align}
Clearly, our previous 2-D DFT expansion \eqref{System_Channel_Det_subs}, \eqref{BEM_dft} is a special case of \eqref{BEM}.

\subsection{1-D and 2-D Basis Expansions}\label{sec:BEM}

We will choose a basis $\{ v_{m,i}[\lambda,\kappa] \}$ that is adapted to the channel model (\ref{dominant_impresp_cont})
(but not to the specific channel parameters $P$, $\eta_p$, $\tau_p$, and $\nu_p$ in (\ref{dominant_impresp_cont})).
Equation (\ref{dominant_impresp_cont}) suggests that the coefficients $\beta_{m,i}$ should be sparse for the elementary single-scatterer channel
$h^{(\tau_1,\nu_1)}(t,\tau) \triangleq \delta(\tau\!-\!\tau_1) \, e^{\jmath 2\pi \nu_1 t }$, for all $\tau_1 \in [0, \tau_{\max}]$ and $\nu_1 \in [-\nu_{\max},\nu_{\max}]$.
Specializing \eqref{Spread_fct_dom} to $P=1$ and $\eta_{1}=1$, and using \eqref{F_S_A},
the 2-D DFT expansion \eqref{System_Channel_Det_subs} yields after a straightforward calculation
\be
\label{exp_sinc}
    H_{\lambda\,\Delta L,\,\kappa\,\Delta K} \eq\rmv \sum_{m=0}^{D-1} \phi^{(\nu_{1})} \Big( m -\rmv \frac{\tau_1}{T_{\rmv{\rm s}}} \Big) \ist
    C^{(\nu_1)}[m,\lambda] \, e^{-\jmath 2 \pi  \frac{\kappa m}{D}} .
\ee
Here, we have set
\be
\label{B_expr}
C^{(\nu_1)}[m,\lambda] \,\triangleq
\sum_{i=-J/2}^{J/2-1} \!\tilde{\alpha}^{(\nu_1)}_{m,i} \,
\frac{1}{\sqrt{J}} \, e^{\jmath 2 \pi \frac{\lambda i}{J}} \ist ,
\ee
where
\be
\label{tilde-alpha-def}
\tilde{\alpha}^{(\nu_1)}_{m,i} \,\triangleq\, \sqrt{J} \sum_{q=0}^{N-1} \psi^{(\nu_1)} [i+qL] \,\ist A^{*}_{\gamma,g}\rmv \bigg( m, \frac{i + qL}{N_{r}} \bigg) \,,
\ee
with $\psi^{(\nu_1)}[i] \triangleq e^{\jmath \pi (\nu_1 T_{\rmv{\rm s}} - i/N_{r}) (N_{r}-1) } \, \psi(i-\rmv \nu_1 T_{\rmv{\rm s}} N_{r} )$.

According to \eqref{tilde-alpha-def}, the poor decay of $\psi( x )$ entails a poor decay of $\tilde{\alpha}^{(\nu_1)}_{m,i}$ with respect to $i$.
To improve the decay, we replace the 1-D DFT \eqref{B_expr} by a general 1-D basis expansion
\begin{align}
  C^{(\nu_1)}[m,\lambda] &\eq\! \sum_{i=-J/2}^{J/2-1} \!\tilde{\beta}^{(\nu_1)}_{m,i} \ist b_{m,i}[\lambda] \,, \quad\; \color{white}{m=0,\dots,D\!-\!1\,,} \nonumber\\
   & \hspace*{5mm} m=0,\dots,D\!-\!1\,, \;\;  \lambda=0,\ldots,J\!-\!1\,,
\label{exp_v}
\end{align}
with a family of bases ${\{b_{m,i}[\lambda]\}}_{i=-J/2,\dots,J/2-1}$, $m = 0,\dots,D\!-\!1$ that are orthonormal (i.e.,
$\sum_{\lambda=0}^{J-1}b_{m,i_{1}}[\lambda] \, b_{m,i_{2}}^{*}[\lambda]=\delta[i_{1} \!-i_{2}]$ for all $m$) and do not depend on the
value of $\nu_1$ in $C^{(\nu_1)}[m,\lambda]$. The idea is to choose the 1-D bases
${\{b_{m,i}[\lambda]\}}_{i=-J/2,\dots,J/2-1}$ such that the coefficient
vector $\big[\tilde{\beta}^{(\nu_1)}_{m,-J/2} \cdots\ist
\tilde{\beta}^{(\nu_1)}_{m,J/2-1}\big]^{T}\!$ is sparse for all
$m$ and all $\nu_1\in [-\nu_{\max},\nu_{\max}]$. Substituting
(\ref{exp_v}) back into (\ref{exp_sinc}), we obtain
\begin{align*}
  &H_{\lambda\,\Delta L,\,\kappa\,\Delta K} \eq\rmv\sum_{m=0}^{D-1} \sum_{i=-J/2}^{J/2-1} \!\phi^{(\nu_{1})} \Big( m -\rmv \frac{\tau_1}{T_{\rmv{\rm s}}}
    \Big) \, \tilde{\beta}^{(\nu_1)}_{m,i} \\
   & \hspace*{48mm}  \times b_{m,i}[\lambda] \, e^{-\jmath 2 \pi \frac{\kappa m}{D}} .
\end{align*}
This can now be identified with the 2-D basis expansion \eqref{BEM}, with the orthonormal 2-D basis
\be
\label{Basis_functions}
v_{m,i}[\lambda,\kappa ] \,\triangleq\ist \frac{1}{\sqrt{D}} \,\ist b_{m,i}[\lambda] \, e^{-\jmath 2 \pi \frac{\kappa m}{D}}
\ee
and the 2-D
coefficients $\beta_{m,i}^{(\tau_1,\nu_1)} \triangleq \sqrt{D} \,\ist\phi^{(\nu_{1})} \big( m - \frac{\tau_1}{T_{\rmv{\rm s}}} \big) \, \tilde{\beta}^{(\nu_1)}_{m,i}$.
The basis functions $v_{m,i}[\lambda,\kappa]$ are seen to agree with our previous 2-D DFT basis functions
$u_{m,i}[\lambda,\kappa] = (1/\sqrt{JD}) \, e^{-\jmath 2 \pi ( \kappa m/D - \lambda i/J )}$
with respect to $\kappa$, but they are different with respect to $\lambda$ because $(1/\sqrt{J}) \,  e^{\jmath 2 \pi \lambda i/J}$ is replaced by $b_{m,i}[\lambda]$.
Furthermore, the sparsity of $\beta_{m,i}^{(\tau_1,\nu_1)}$ in the $i$ direction is governed by the new 1-D coefficients
$\tilde{\beta}^{(\nu_1)}_{m,i}$,
which are potentially sparser than the previous 1-D coefficients $\tilde{\alpha}^{(\nu_1)}_{m,i}$ in \eqref{B_expr} that were based on the 1-D DFT basis
$\big\{ (1/\sqrt{J}) \,  e^{\jmath 2 \pi \lambda i/J} \big\}$.

These considerations can be immediately extended to the
multiple-scatterer case. When the channel comprises $P$ scatterers
as in \eqref{dominant_impresp_cont}, the coefficients are
$\beta_{m,i} = \sum_{p=1}^{P} \eta_{p}
\,\beta_{m,i}^{(\tau_{p},\nu_{p})}\!$. If each coefficient
sequence $\beta_{m,i}^{(\tau_{p},\nu_{p})}\!$ is $S$-sparse,
$\beta_{m,i}$ is $PS$-sparse. Note that, by construction, our
basis $\{ v_{m,i}[\lambda,\kappa] \}$ does not depend on the
channel parameters $P$, $\eta_p$, $\tau_p$, and $\nu_p$, and its
formulation is not explicitly based on the channel model
\eqref{dominant_impresp_cont}. The use of the generalized 2-D
basis $\{ v_{m,i}[\lambda,\kappa] \}$ in \eqref{Basis_functions}
comes at the cost of an increased computational complexity,
because efficient FFT algorithms can only be applied with respect
to $\kappa$ but not with respect to $\lambda$. However, if $J$ is
not too large, the additional complexity is small. Optimal designs
of the 1-D bases ${\{b_{m,i}[\lambda]\}}_{i=-J/2,\dots,J/2-1}$
will be presented in Section \ref{sec:basis_opt}.

\subsection{Generalized Compressive Channel Estimator}\label{sec:chest_basis}

A CS-based channel estimation scheme that uses the generalized basis expansion \eqref{BEM}
can be developed similarly as in Section \ref{sec:chest}. We can write \eqref{BEM} as (cf.\ \eqref{ch_est})
$\mathbf{h} = \mathbf{V} \bbeta$, with a unitary matrix $\mathbf{V}$. Here, $\bbeta$ and $\mathbf{V}$ are defined in an analogous manner as, respectively, $\aalpha$ and $\mathbf{U}$ were defined in Section \ref{sec:chest}. Reducing this relation to the pilot positions yields (cf.\ \eqref{ch_est_pilot_1})
$\mathbf{h}^{({\rm p})} = \mathbf{V}^{({\rm p})} \bbeta = \mathbf{\Phi} \mathbf{x}$, with $\mathbf{\Phi} \triangleq \mathbf{V}^{({\rm p})} \mathbf{D}$
and $\mathbf{x} \triangleq \mathbf{D}^{-1} \bbeta$, where the diagonal matrix $\mathbf{D}$ is chosen such that all columns of $\mathbf{\Phi}$ have unit $\ell_2$-norm.
Finally, we replace the unknown vector $\mathbf{h}^{({\rm p})}$ by
its pilot-based estimate, again denoted as $\mathbf{y}$. Using
\eqref{pilot_ch_est}, we then obtain the \emph{measurement equation} (cf.\ \eqref{chest_sparse_reconstr})
$\mathbf{y}= \mathbf{\Phi} \mathbf{x} + \mathbf{z}$, where $\mathbf{z}$ is again the vector with entries $\tilde z_{l,k}/p_{l,k} \big|_{(l,k) \in
\mathcal{P}}\ist$. As in Section \ref{sec:chest}, our task is to recover the length-$JD$ vector $\mathbf{x}$ from the known
length-$|\mathcal{P}|$ vector $\mathbf{y}$, based on the measurement equation.
From the resulting estimate of $\mathbf{x}$, estimates of the channel coefficients $H_{l,k}$ on
the subsampled grid $\mathcal{G}$ are obtained via \eqref{BEM} by means of the equivalence of $\beta_{m,i}$ and $\bbeta =\mathbf{D} \mathbf{x}$.
Inverting\footnote{Note
that the 1-D part of \eqref{BEM} corresponding to index $m$ equals the respective 1-D part of \eqref{System_Channel_Det_subs} (1-D DFT),
since $v_{m,i}[\lambda,\kappa ] = (1/\sqrt{D}) \,\ist b_{m,i}[\lambda] \, e^{-\jmath 2 \pi \kappa m/D}$. Hence,
the transformation \eqref{BEM} and the
inverted transformation \eqref{System_Channel_Det_subs} have to be applied only with respect to the index
$i$.}
\eqref{System_Channel_Det_subs} and applying \eqref{System_Channel_Det_1} then yields estimates of all channel coefficients $H_{l,k}$. As discussed further above, we
can expect $\bbeta$ and, in turn, $\mathbf{x}$ to be approximately
sparse provided the 1-D bases $\{b_{m,i}[\lambda] \}$ are chosen
appropriately. Hence, our channel estimation problem is again recognized to be a
sparse reconstruction problem of the form \eqref{sparse_reconstr},
with dimensions $M \rmv=\rmv \text{dim} \{ \mathbf{x} \} \rmv=\rmv
JD$ and $Q \rmv=\rmv \text{dim} \{ \mathbf{y} \} \rmv=\rmv |\mathcal{P}|$. We can thus use a CS recovery technique
to obtain an estimate of $\mathbf{x}$.

For consistency with the CS framework of Section \ref{sec:chest}, we select the pilot positions
uniformly at random within the subsampled time-frequency grid $\mathcal{G}$.
For BP and CoSaMP, to achieve a small upper bound on the reconstruction error, the
number of pilots should satisfy condition (\ref{UUP_satisfy_suffcond}), i.e.,
\[
    |\mathcal{P}| \,\geq\, C \, \gamma^{-2} \ist \big( \rmv\ln (JD) \big)^{4} \ist \mu_\mathbf{V}^{2} \ist S \ist \ln(1/\eta) \,,
\]
where $S$ is the sparsity of $\mathbf{x}$ and $\mu_\mathbf{V}$ is the coherence of $\mathbf{V}$.
Note that $S$ depends on the chosen basis $\{ v_{m,i}[\lambda,\kappa] \}$;
furthermore, $\mu_\mathbf{V} \!\ge\! 1$ (for the DFT basis, we had $\mu_\mathbf{U} \!=\! 1$).
Thus, the performance gain due to the better sparsity may be reduced to a certain extent because of the larger coherence.

\section{Basis Optimization}\label{sec:basis_opt}

We now discuss the optimal design of the 1-D bases ${\{b_{m,i}[\lambda]\}}$.

\subsection{Basis Optimization Framework}\label{sec:basis_opt_framework}

The orthonormal 1-D bases ${\{b_{m,i}[\lambda]\}}_{i = -J/2,\dots,J/2-1}$, $m=0,\dots,D\!-\!1$ should be such that the
coefficient vectors $\big[\tilde{\beta}^{(\nu)}_{m,-J/2}
\cdots\ist \tilde{\beta}^{(\nu)}_{m,J/2-1}\big]^{T}\!$ are sparse
for all $m$ and all $\nu\in [-\nu_{\max},\nu_{\max}]$ (the maximum
Doppler frequency shift $\nu_{\max}$ is assumed known).
For our optimization, we slightly relax this requirement in that we only require a sparse coefficient vector
for a finite number of uniformly spaced Doppler frequencies $\nu \!\in\! \mathcal{D}$, where
$\mathcal{D} \ist\triangleq \big\{\nu_{\Delta}  d,\,\, d=-\lceil\nu_{\max}/\nu_{\Delta}\rceil,\dots, \lceil \nu_{\max}/\nu_{\Delta}\rceil \big\}$
with some Doppler frequency spacing $\nu_{\Delta}$.

Regarding the choice of $\nu_{\Delta}$, it is interesting to note that for
the ``canonical spacing'' given by $\nu_{\Delta}=1/(T_{\rmv{\rm s}} N_{r})$, the coefficients $\tilde{\alpha}^{(\nu_{\Delta} d)}_{m,i}$
in the 1-D DFT expansion \eqref{B_expr} are $1$-sparse with respect to $i$. Indeed,
$\psi^{(\nu_1)}[i] = e^{\jmath \pi (\nu_1 T_{\rmv{\rm s}} - i/N_{r}) (N_{r}-1) } \, \psi(i-\rmv \nu_1 T_{\rmv{\rm s}} N_{r} )$
here simplifies to
$\psi^{(\nu_{\Delta}  d)}[i] = e^{\jmath \pi (d-i) (N_{r}-1)/N_{r} } \ist \psi (i \rmv-\rmv d) = \delta_{N_{r}}[i \rmv-\rmv d]$,
where $\delta_{N_{r}}[i]$ is the $N_{r}$-periodic unit sample (i.e., $\delta_{N_{r}}[i]$ is $1$ if $i$ is a multiple of $N_{r}$ and $0$ otherwise).
Expression \eqref{tilde-alpha-def} then reduces to
\begin{align*}
  \tilde{\alpha}^{(\nu_{\Delta} d)}_{m,i} &\eq \sqrt{J} \sum_{q=0}^{N-1} \delta_{N_{r}}[i \rmv-\rmv d + qL] \,
  A^{*}_{\gamma,g}\rmv \bigg( m, \frac{i + qL}{N_{r}} \bigg) \\
   & \eq \delta_{N_{r}}\big[i \rmv-\rmv \tilde{d}\ist \big] \ist A^{*}_{\gamma,g}\rmv \bigg( m, \frac{d}{N_{r}}
  \bigg)\,,
\end{align*}
where $\tilde{d}$ depends on $d$ but not on $i$. Thus, for
$\nu_{\Delta}=1/(T_{\rmv{\rm s}} N_{r})$, the coefficients
obtained using the 1-D DFT basis $\big\{ b_{m,i}[\lambda] =
(1/\sqrt{J}) \,  e^{\jmath 2 \pi \lambda i/J} \big\}$ are
$1$-sparse (no leakage effect). This means that the 1-D DFT basis
would be optimal; no other basis could do better. We therefore
choose a Doppler spacing that is twice as dense, i.e.,
$\nu_{\Delta}=1/(2 \ist T_{\rmv{\rm s}} N_{r})$. That is, we
define $\mathcal{D}$ such that it includes also the Doppler
frequencies located midway between any two adjacent canonical
sampling points. For these frequencies---given by $\nu_{\Delta} d$
for odd $d$---the leakage (obtained with the DFT basis) is
maximal.

Because the basis $\{b_{m,i}[\lambda]\}$ is orthonormal, the expansion coefficients $\tilde{\beta}^{(\nu)}_{m,i}$
defined by \eqref{exp_v} can be calculated as the inner products
$\tilde{\beta}^{(\nu)}_{m,i} = \sum_{\lambda=0}^{J-1} C^{(\nu)}[m,\lambda] \, b_{m,i}^*[\lambda]$, $i=-J/2,\dots,J/2\!-\!1$. This can be rewritten as
\[
    \tilde{\bbeta}^{(\nu)}_{m} \rmv\eq \mathbf{B}_{m} \ist \mathbf{c}^{(\nu)}_{m} \ist ,
\]
with the length-$J$ vectors $\tilde{\bbeta}^{(\nu)}_{m}
\rmv\triangleq \big[\tilde{\beta}^{(\nu)}_{m,-J/2} \cdots\ist
\tilde{\beta}^{(\nu)}_{m,J/2-1} \big]^{T}\!$ and
$\mathbf{c}^{(\nu)}_{m} \rmv\triangleq \big[ C^{(\nu)}[m,0]
\ist\cdots\ist C^{(\nu)}[m,J-1]\big]^{T}\!$ and the unitary $J
\!\times\! J$ matrix $\mathbf{B}_{m}$ with entries
$(\mathbf{B}_{m})_{i+1,\lambda+1} = b_{m,i-J/2}^*[\lambda]$. We
can now state the basis optimization problem as follows. \emph{For
given vectors $\mathbf{c}^{(\nu)}_{m}\!$, $m=0,\dots,D\!-\!1$,
with $\mathbf{c}^{(\nu)}_{m}$ defined as described above, find
unitary $J \!\times\! J\rmv$ matrices $\mathbf{B}_{m}\!$ not
dependent on $\nu$  such that the vectors
$\ist\tilde{\bbeta}^{(\nu)}_{m} \rmv= \mathbf{B}_{m}  \ist
\mathbf{c}^{(\nu)}_{m}\rmv$ are maximally sparse for all} $\nu
\!\in\rmv \mathcal{D}$.

For the sake of algorithmic simplicity, we will measure
the sparsity of $\tilde{\bbeta}^{(\nu)}_{m}$ by the $\ell_1$-norm
or, more precisely, by the $\ell_{1}$-norm averaged over all $\nu
\rmv\in\rmv \mathcal{D}$, i.e., $\frac{1}{|\mathcal{D}|}
\sum_{\nu\in \mathcal{D}} \big\|\tilde{\bbeta}^{(\nu)}_{m}
\big\|_{1} = \frac{1}{|\mathcal{D}|} \sum_{\nu\in \mathcal{D}}
\big\|\mathbf{B}_{m}  \ist \mathbf{c}^{(\nu)}_{m} \big\|_{1}$.
Thus, our basis optimization problem is formulated as the $D$
constrained minimization problems\footnote{
We
note that the optimization problem (\ref{opt1}) is similar to {\em dictionary learning} problems that have recently been considered
in \cite{ahelbr06,krmura03,Gribonval08}. In \cite{Gribonval08},
conditions for the local identifiability of orthonormal bases by
means of $\ell_{1}$ minimization have been derived. An $\ell_0$-norm based sparsity-enhancing
basis design has been proposed in the MIMO context in \cite{schniter_sayeed04}.
Furthermore, basis adaptation and selection at the receiver has been considered in the ultrawideband context in \cite{WangArce_ICUWB08}.
}
\be \label{opt1}
    \hat{\mathbf{B}}_{m} \eq\ist \arg \hspace*{.4mm}\min_{\hspace*{-6.5mm}\rule{0mm}{2.6mm}\mathbf{B}_{m} \in\, \mathcal{U}} \ist \sum_{\nu\in \mathcal{D}}
    \big\|\mathbf{B}_{m}  \ist \mathbf{c}^{(\nu)}_{m} \big\|_{1} \,, \;\;\; m=0,\dots,D\!-\!1 \,,
\ee where $\mathcal{U}$ denotes the set of all unitary $J
\!\times\! J$ matrices. Note that the vectors
$\mathbf{c}^{(\nu)}_{m}\!$ are known because they follow from the
function $C^{(\nu)}[m,\lambda]$, which is given by (see
\eqref{B_expr}, \eqref{tilde-alpha-def}) $C^{(\nu)}[m,\lambda] =
\sum_{i=-J/2}^{J/2-1} \sum_{q=0}^{N-1} \psi^{(\nu)}[i +qL] \,
A^{*}_{\gamma,g}\rmv \big( m, \frac{i +qL}{N_{r}} \big) \ist
e^{\jmath 2 \pi \lambda i/J}$. It is seen that the optimal bases
characterized by the matrices $\hat{\mathbf{B}}_{m}$ depend on
$N$, $L$, $J$, $g[n]$, $\gamma[n]$, and (via the definition of
$\mathcal{D}$) $\nu_{\max}$, but not on any other channel
properties.

For classical CP-OFDM with CP length $N\!-\!K \geq D\!-\!1$, we
have $A_{\gamma,g}( m,\xi)=A_{\gamma,g}( 0, \xi)$ for all
$m=1,\dots,D\!-\!1$, so $C^{(\nu)}[m,\lambda] =
C^{(\nu)}[0,\lambda]$ (see \eqref{B_expr},
\eqref{tilde-alpha-def}) and thus $\mathbf{c}^{(\nu)}_{m} \!\rmv=
\mathbf{c}^{(\nu)}_{0}\!$. Because $\mathbf{c}^{(\nu)}_{m}$ no
longer depends on $m$, only one basis $\mathbf{B}$ (instead of $D$
different bases $\mathbf{B}_m$, $m=0,\dots,D\!-\!1$) has to be
optimized.

\subsection{Statistical Basis Optimization}\label{sec:basis_stat_opt}

The basis optimization framework presented above can be extended
to take into account prior statistical information about the
channel. Let us again consider the single-scatterer channel
$h^{(\tau_1,\nu_1,\eta_{1})}(t,\tau) = \eta_{1} \ist
\delta(\tau\!-\!\tau_1) \, e^{\jmath 2\pi \nu_1 t }$, now
including a path gain $\eta_{1}$. We assume that $\tau_1$,
$\nu_1$, and $\eta_{1}$ are random, with $(\tau_1,\nu_1)$
distributed according to a known probability density function
(pdf) $p(\tau_1,\nu_1)$, and $\eta_{1}$ given $(\tau_1,\nu_1)$
being zero-mean, circularly symmetric complex Gaussian with known
variance $\sigma^2(\tau_1,\nu_1)$. As before, we consider a 2-D
expansion of the subsampled channel coefficients
$H_{\lambda\,\Delta L,\kappa\,\Delta K}$ into (deterministic)
orthonormal basis functions $v_{m,i}[\lambda,\kappa]$, i.e.,
$H_{\lambda\,\Delta L,\kappa\,\Delta K} =
\sum_{m=0}^{D-1}\sum_{i=-J/2}^{J/2-1} \beta_{m,i} \,
v_{m,i}[\lambda,\kappa]$, $\lambda =0,\ldots,J\!-\!1$,
$\kappa=0,\ldots, D\!-\!1$. Clearly, the vector $\bbeta$ of
expansion coefficients $\beta_{m,i}$ (which is defined as in
Section \ref{sec:chest_basis}) now is a random vector. Our goal is
to find basis functions $v_{m,i}[\lambda,\kappa]$ (or,
equivalently, a unitary matrix $\mathbf{V}$, defined as in Section
\ref{sec:chest_basis}) such that $\bbeta = \bbeta(\mathbf{V})$ is
maximally sparse \emph{on average}. Measuring the sparsity of
$\bbeta$ by the $\ell_1$-norm for convenience, we obtain the
optimization problem \be \label{stat_opt}
    \hat{\mathbf{V}} \eq\ist \arg \hspace*{.4mm}\min_{\hspace*{-6.5mm}\rule{0mm}{2.6mm}\mathbf{V} \in\, \mathcal{U}'} \,
    {\rm E}\big\{\big\|\bbeta(\mathbf{V}) \big\|_{1} \big\} \,,
\ee where ${\rm E}\{\cdot\}$ denotes expectation and
$\mathcal{U}'\rmv$ denotes the set of all unitary $JD \!\times\!
JD$ matrices.

Again, we set $v_{m,i}[\lambda,\kappa] \triangleq (1/\sqrt{D}) \,\ist b_{m,i}[\lambda] \, e^{-\jmath 2 \pi \kappa m/D}$
with a family of orthonormal 1-D bases $\{ b_{m,i}[\lambda] \}$. Then, \eqref{stat_opt} reduces to the minimization of
${\rm E}\big\{\big\|\bbeta \big\|_{1} \big\}$ with respect to $\{ b_{m,i}[\lambda] \}$.
For the single-scatterer channel, the $\ell_{1}$-norm of $\bbeta = \bbeta^{(\tau_1,\nu_1,\eta_{1})}$ can be shown to be
\begin{align*}
  \big\|\bbeta^{(\tau_1,\nu_1,\eta_{1})} \big\|_{1}
  &\rmv\eq \sqrt{D} \, |\eta_{1}|
  \sum_{m=0}^{D-1} \Big|\phi^{(\nu_{1})} \Big( m -\rmv \frac{\tau_1}{T_{\rmv{\rm s}}}\Big)\Big| \nonumber\\
   & \hspace*{15mm} \times \sum_{i=-J/2}^{J/2-1}\left|\sum_{\lambda=0}^{J-1} C^{(\nu_{1})}[m,\lambda] \ist b_{m,i}^{*}[\lambda]\right| \ist,
\end{align*}
with $C^{(\nu_1)}[m,\lambda]$ as in \eqref{B_expr}, \eqref{tilde-alpha-def}. We note that $|\eta_{1}|$ given $(\tau_1,\nu_1)$ is Rayleigh distributed with mean
$\sigma(\tau_1,\nu_1)\sqrt{\pi/2}$. Hence, ${\rm E}\big\{ \big\|\bbeta^{(\tau_1,\nu_1,\eta_{1})} \big\|_{1}\big\}$
is given by (hereafter, we write $\tau,\nu,\eta$ instead of $\tau_1,\nu_1,\eta_{1}$)
\begin{align}
  {\rm E}\big\{ \big\|\bbeta^{(\tau,\nu,\eta)}\big\|_{1}\big\} &\eq \sqrt{\frac{D\pi}{2}} \rmv
  \int_{-\infty}^\infty \ist \sum_{m=0}^{D-1}G^{(\nu)}[m] \nonumber\\
   & \hspace*{5mm} \times \sum_{i=-J/2}^{J/2-1}\left|\sum_{\lambda=0}^{J-1} C^{(\nu)}[m,\lambda] \ist b_{m,i}^{*}[\lambda]\right|  \ist d \nu \,,
\label{fprime_sing_scat_exp}
\end{align}
with
\[
G^{(\nu)}[m] \,\triangleq \int_{-\infty}^\infty \!\sigma(\tau,\nu)
\, \Big|\phi^{(\nu)} \Big(m -\rmv \frac{\tau}{T_{\rmv{\rm
s}}}\Big)\Big|
  \,\ist p(\tau,\nu) \, d \tau \,>\, 0 \,.
\]
It follows that minimizing \eqref{fprime_sing_scat_exp}
with respect to $\{ b_{m,i}[\lambda] \}$ amounts to minimizing
\be \label{min_prob}
    \int_{-\infty}^\infty \sum_{i=-J/2}^{J/2-1}\left|\sum_{\lambda=0}^{J-1} C^{(\nu)}[m,\lambda] \ist b_{m,i}^{*}[\lambda]\right|
     G^{(\nu)}[m] \, d \nu
\ee
for all $m=0,\ldots,D\!-\!1$. Note that $G^{(\nu)}[m]$ can be computed from the known statistics. In vector-matrix notation, with
$\mathbf{c}^{(\nu)}_{m} \rmv\triangleq \big[ C^{(\nu)}[m,0] \ist\cdots\, C^{(\nu)}[m,J-1] \ist \big]^{T}\!$ and the unitary
$J \!\times\! J$ matrix $\mathbf{B}_{m}$ with entries $(\mathbf{B}_{m})_{i+1,\lambda+1} \triangleq b_{m,i-J/2}^*[\lambda]$, minimization of \eqref{min_prob}
can be equivalently written as minimization of
\be
\label{opt_stat}
   \int_{-\infty}^\infty \big\|\mathbf{B}_{m} \mathbf{c}^{(\nu)}_{m} \big\|_{1} \ist G^{(\nu)}[m]  \, d \nu
\ee over the set $\mathcal{U}$ of all unitary $J \!\times\! J$
matrices $\mathbf{B}_{m}$, for $m=0,\dots,D\!-\!1$. Approximating
this integral by its Riemannian sum\footnote{Alternatively, the
integral can be interpreted as an expectation with respect to
$\nu$ and computed by means of Monte Carlo techniques. This is
especially advantageous if the maximum Doppler frequency is
unknown.} over the set $\mathcal{D} \ist\triangleq
\big\{\nu_{\Delta}  d,\,\,
d=-\lceil\nu_{\max}/\nu_{\Delta}\rceil,\dots, \lceil
\nu_{\max}/\nu_{\Delta}\rceil \big\}$ with $\nu_{\Delta}=1/(2 \ist
T_{\rmv{\rm s}} N_{r})$, for a given maximum Doppler frequency
$\nu_{\max}$, the minimization problem can be finally stated as
\be \label{opt_stat_final}
    \hat{\mathbf{B}}_{m} \rmv\eq\ist \arg \hspace*{.4mm}\min_{\hspace*{-6.5mm}\rule{0mm}{2.6mm}\mathbf{B}_{m} \in\, \mathcal{U}} \ist
    \sum_{\nu \in \mathcal{D}} \big\|\mathbf{B}_{m} \tilde{\mathbf{c}}^{(\nu)}_{m} \big\|_{1} \,, \;\;
    \text{with} \;\, \tilde{\mathbf{c}}^{(\nu)}_{m} \ist\triangleq\, \mathbf{c}^{(\nu)}_{m} \ist G^{(\nu)}[m]  \,,
\ee
for $m=0,\dots,D\!-\!1$. This is recognized to be of the same form as \eqref{opt1}.

In practice, the channel statistics $p(\tau,\nu)$,
$\sigma^2(\tau,\nu)$ will deviate from the true statistics to some
extent, so that the basis matrices $\hat{\mathbf{B}}_{m}$ obtained
as described above will be different from the truly optimal ones.
An interesting question is as to how this difference affects the
average sparsity of the expansion coefficient vector
$\bbeta^{(\tau,\nu,\eta)}\rmv$. For simplicity, we measure the
average sparsity by ${\rm E}\{ {\|\bbeta \|}_{1} \}$, and we
assume that the optimization criterion is minimization of
\eqref{opt_stat} (which, after all, is almost equivalent to
\eqref{opt_stat_final}) and, further, that $\Delta L \rmv=\rmv 1$
or equivalently $J \rmv=\rmv L$ (i.e., no subsampling with respect
to $l$). Let $\bbeta$ and $\tilde\bbeta$ denote the expansion
coefficient vectors obtained for the true and incorrect bases,
respectively. Then, one can show the following bound on the
normalized difference of the average sparsities of $\tilde\bbeta$
and $\bbeta$:
\begin{align*}
  &\frac{ \big| {\rm E}\{ {\|\tilde\bbeta \|}_{1} \} - {\rm E}\{
{\|\bbeta \|}_{1} \} \big|}{ {\rm E}\{ {\|\bbeta \|}_{1} \} } \\[1mm]
   & \hspace*{1mm} \leq\, 2 \sqrt{L} \; \frac{ \int_{-\infty}^\infty
\sum_{m=0}^{D-1} \big| \tilde{G}^{(\nu)}[m] - G^{(\nu)}[m] \big|
\,
  \big| A_{\gamma,g}( m, \nu T_{\rmv \text{s}}) \big| \, d\nu }{
  \int_{-\infty}^\infty \sum_{m=0}^{D-1} G^{(\nu)}[m] \big| A_{\gamma,g}( m, \nu T_{\rmv \text{s}}) \big| \, d\nu } \,,
\end{align*}
where $\tilde{G}^{(\nu)}[m]$ is defined analogously to $G^{(\nu)}[m]$ but with the incorrect statistics.

\subsection{Basis Optimization Algorithm}\label{sec:opt_algo}

Because the minimization problems \eqref{opt1} and
\eqref{opt_stat_final} are nonconvex (since $\mathcal{U}$ is not a
convex set), standard convex optimization techniques cannot be
used. We therefore propose an approximate iterative algorithm that
relies on the following facts \cite{golub96}. (i) Every unitary $J
\!\times\! J$ matrix $\mathbf{B}$ can be represented in terms of a
Hermitian $J \!\times\! J$ matrix $\mathbf{A}$ as $\mathbf{B} =
e^{\jmath \mathbf{A}}$. (ii) The matrix exponential $\mathbf{B} =
e^{\jmath \mathbf{A}}$ can be approximated by its first-order
Taylor expansion, i.e., $\mathbf{B} \approx \mathbf{I}_{J} +
\jmath \mathbf{A}$, where $\mathbf{I}_{J}$ is the $J \!\times\! J$
identity matrix. Even though $\mathbf{B}$ is unitary and
$\mathbf{I}_{J} + \jmath \mathbf{A}$ is not, this approximation
will be good if ${\|\mathbf{A}\|}_\infty$ is small, where
${\|\mathbf{A}\|}_\infty$ denotes the largest modulus of all
entries of $\mathbf{A}$. Because of this condition, we construct
$\mathbf{B}_m$ iteratively: starting with the DFT basis, we
perform a \emph{small} update at each iteration, using the
approximation $\mathbf{B} \approx \mathbf{I}_{J} + \jmath
\mathbf{A}$ in the optimization criterion \emph{but not for
actually updating} $\mathbf{B}_m$ (thus, the iterated
$\mathbf{B}_m$ is always unitary). More specifically, at the
$\iter$th iteration, we consider the following update of the
unitary matrix $\mathbf{B}_m^{(\iter)}$:
\[
\mathbf{B}_m^{(\iter+1)} \rmv\eq e^{\jmath \mathbf{A}_m^{\!(\iter)}} \ist \mathbf{B}_m^{(\iter)} \,,
\]
where $\mathbf{A}_m^{\!(\iter)}$ is a  small Hermitian matrix that remains to be optimized. Note that $\mathbf{B}_m^{(\iter+1)}$ is again unitary
because both $\mathbf{B}_m^{(\iter)}$ and $e^{\jmath \mathbf{A}_m^{\!(\iter)}}$ are unitary.

Ideally, we would like to optimize $\mathbf{A}_m^{\!(\iter)}$
according to \eqref{opt1} (or \eqref{opt_stat_final}), i.e., by minimizing $\sum_{\nu\in
\mathcal{D}} \big\| \mathbf{B}_m^{(\iter+1)} \ist
\mathbf{c}_m^{(\nu)} \big\|_{1} = \sum_{\nu\in \mathcal{D}}
\big\|e^{\jmath \mathbf{A}_m^{\!(\iter)}} \ist
\mathbf{B}_m^{(\iter)} \ist \mathbf{c}_m^{(\nu)} \big\|_{1}$.
Since this problem is still nonconvex, we use the approximation $e^{ \jmath \mathbf{A}} \approx \mathbf{I}_{J} + \jmath \mathbf{A}$, and thus the final minimization
problem at the $\iter$th iteration is
\be \label{opt2}
    \hat{\mathbf{A}}_m^{\!(\iter)}  \eq \arg \hspace*{.4mm}\min_{\hspace*{-5.8mm}\rule{0mm}{2.9mm}\mathbf{A} \in \mathcal{A}_{\iter}} \ist
    \sum_{\nu\in \mathcal{D}} \big\| (\mathbf{I}_{J} + \jmath \mathbf{A} ) \ist \mathbf{B}_m^{(\iter)} \mathbf{c}_m^{(\nu)} \big\|_{1} \,.
\ee Here, $\mathcal{A}_{\iter}$ is the set of all Hermitian $J
\!\times\! J$  matrices $\mathbf{A}$ that are small in the sense
that ${\|\mathbf{A}\|}_\infty \leq \rho_{\iter}$, where
$\rho_{\iter}$ is a positive constraint level (a small
$\rho_{\iter}$ ensures a good accuracy of our approximation
$\mathbf{B} \approx \mathbf{I}_{J} + \jmath \mathbf{A}$ and also
that $e^{ \jmath \hat{\mathbf{A}}_m^{\!(\iter)} }$ is close to
$\mathbf{I}_{J}$). The problem (\ref{opt2}) is convex and thus can
be solved by standard convex optimization techniques
\cite{boyd_conv_opt01}.

The next step at the $\iter$th iteration is to test whether the
cost function is smaller for the new unitary matrix $e^{\jmath
\hat{\mathbf{A}}_m^{\!(\iter)} } \ist \mathbf{B}_m^{(\iter)}\!$,
i.e., whether $\sum_{\nu\in \mathcal{D}} \big\| e^{ \jmath
\hat{\mathbf{A}}_m^{\!(\iter)} } \ist \mathbf{B}_m^{(\iter)} \ist
\mathbf{c}_m^{(\nu)} \big\|_{1} < \sum_{\nu\in \mathcal{D}}
\big\|\mathbf{B}_m^{(\iter)} \ist \mathbf{c}_m^{(\nu)}
\big\|_{1}$. In the positive case, we actually perform the update
of $\mathbf{B}_m^{(\iter)}$ and we retain the constraint level
$\rho_{\iter}$ for the next iteration, i.e.,
\[
\mathbf{B}_m^{(\iter+1)} =\, e^{\jmath \hat{\mathbf{A}}_m^{\!(\iter)} } \ist \mathbf{B}_m^{(\iter)} \,, \qquad\quad
\rho_{\iter+1} \ist=\ist \rho_{\iter} \,.
\]
Otherwise, we reject the update of $\mathbf{B}_m^{(\iter)}$ and
reduce the constraint level $\rho_{\iter}$, i.e.,
\[
\mathbf{B}_m^{(\iter+1)} =\, \mathbf{B}_m^{(\iter)} \,, \qquad\quad
\rho_{\iter+1} \ist=\ist \frac{\rho_{\iter}}{2} \,.
\]
By this construction, the cost function sequence $\sum_{\nu\in
\mathcal{D}} \big\|\mathbf{B}_m^{(\iter)} $ $\ist
\mathbf{c}_m^{(\nu)} \big\|_{1}$, $r \!=\! 0,1,\ldots$ is
guaranteed to be monotonically decreasing.

The above iteration process is terminated if $\rho_{\iter}$ falls
below a prescribed threshold or if the number of iterations
exceeds a certain value. The iteration process is initialized by
the $J \!\times\! J$ DFT matrix $\mathbf{F}_{\rmv J}$, i.e.,
$\mathbf{B}_m^{(0)} \rmv= \mathbf{F}_{\rmv J}$, because  the DFT
basis was seen in Section \ref{sec:sparsity} to yield a relatively
sparse coefficient vector. We note that efficient algorithms for
computing the matrix exponentials $e^{\jmath
\hat{\mathbf{A}}_m^{\!(\iter)}}$ exist \cite{golub96}. Since the
bases $\{b_{m,i}[\lambda]\}$ (or, equivalently, the basis matrices
$\mathbf{B}_m$) do not depend on the received signal, they have to
be optimized only once before the actual channel estimation
starts.

\begin{figure*}[t]
    \vspace*{-15mm}
\centering
\hspace*{-4mm}\includegraphics[height=6.7cm,width=17cm]{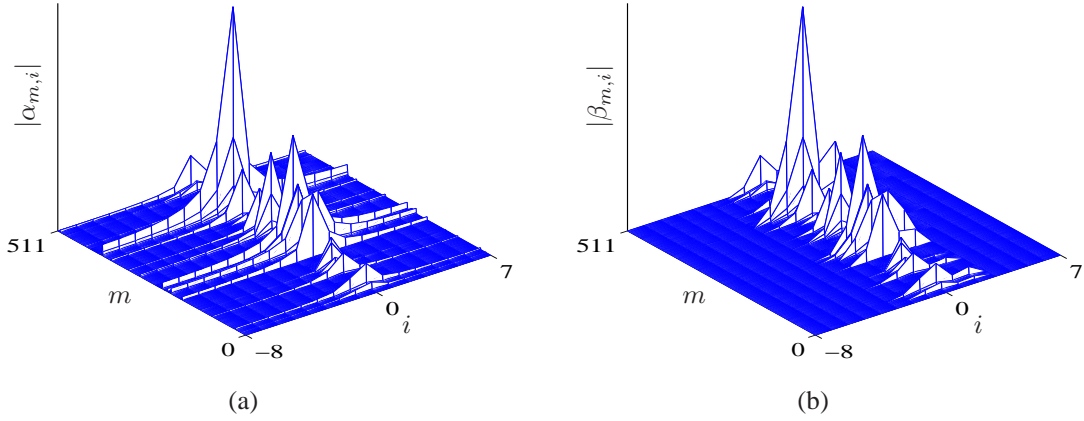}
\put(-402,33){$m$} \put(-292,22){$i$} \put(-187,33){$m$}
\put(-78,22){$i$} \put(-357,-7){(a)} \put(-144,-7){(b)}
\put(-437,99){\begin{sideways}$|\alpha_{m,i}|$\end{sideways}}
\put(-223,99){\begin{sideways}$|\beta_{m,i}|$\end{sideways}}
\vspace*{1mm}
\renewcommand{\baselinestretch}{1.1}\small\normalsize
\caption{Sparsity enhancement obtained with the proposed iterative
basis optimization algorithm: Modulus of the expansion
coefficients for (a) the DFT basis and (b) the optimized basis.}
\label{fig.sparsity} \vspace{1mm}
\end{figure*}

In Fig.\ \ref{fig.sparsity}, we compare the expansion coefficients
$\alpha_{m,i}$ obtained with the DFT basis (see \eqref{BEM_dft})
and $\beta_{m,i}$ obtained with the deterministically optimized
basis (see \eqref{BEM}, \eqref{Basis_functions}) for one channel
realization. The system parameters are as in Sections
\ref{sec:sim_setup} and \ref{sec:sim_BE_det} (first scenario). For
the minimization (\ref{opt2}) (not $m$-dependent, since we
consider a CP-OFDM system), we used the convex optimization
package \texttt{CVX} \cite{Grant_cvx}. It is seen that the basis
optimization yields a significant enhancement of sparsity.

\section{Channel Estimation for Strongly Dispersive Channels}\label{sec:disp_channel}

For strongly dispersive channels, the off-diagonal system channel coefficients (ISI/ICI coefficients)
$\{H_{l,k;l'\!,k'}\}_{(l,k)\neq(l'\!,k')}$ in
\eqref{gen.io.rel_4D} are no longer negligible. Therefore, we now present a compressive channel estimator
that is able to produce reliable estimates of {\it all} channel coefficients $H_{l,k;l'\!,k'}$.

\subsection{Basis Expansion Model}\label{sec:basis_exp_mod}

The proposed channel estimator uses a basis expansion model \cite{leus_eusipco04,Borah99,zemen_sp05} that is different
from the basis expansion considered in Sections \ref{sec:basis_expansion} and \ref{sec:basis_opt}.
The discrete-time channel impulse response $h[n,m]$ is expanded with respect to $n$
into orthonormal basis functions $\ppsi_{i}[n]$, $i=0,\ldots,
N_{r}\!-\!1$, i.e.,
\be \label{h_expans}
h[n,m] \eq\!
\sum_{i=0}^{N_{r}-1} \rmv \T_{h}[m,i] \, \ppsi_{i}[n] \,, \quad
n = 0,\dots,N_{r}\!-\!1 \,,
\ee
with $m$-dependent expansion coefficients
\be \label{spread_fct_gen}
\T_{h}[m,i] \,\triangleq
\sum_{n=0}^{N_{r}-1} \! h[n,m] \, \ppsi^{*}_{i}[n] \,.
\ee
The function $\T_{h}[m,i]$ generalizes the discrete-delay-Doppler
spreading function $S_h[m,i]$ in \eqref{Spread_fct}, which is
reobtained for $\ppsi_{i}[n] = (1/\sqrt{N_{r}}) \, e^{\jmath 2 \pi in/N_{r}}$ (up to a constant factor). Similarly to
\eqref{io_channel_spreading}, the discrete-time channel can now be rewritten as
\begin{align}
  r[n] &\,=\!  \sum_{m=-\infty}^{\infty} \!\sum_{i=0}^{N_{r}-1} \rmv \T_h[m,i] \ist s[n\rmv-\rmv m] \ist \ppsi_{i}[n] \,+\,z[n]
    \,, \nonumber\\[-1mm]
   & \hspace*{44mm} n=0, \ldots,N_{r}\!-\!1 \,.
\label{io_channel_spreading_gen}
\end{align}
We assume that the support of $\T_h[m,i]$ is contained in $\{0,\ldots,D\!-\!1\} \times \{0,\ldots,J\!-\!1\}$
($h[n,m]$ is assumed causal with maximum delay at most $D\!-\!1$).
Combining \eqref{rec-symbol}, \eqref{io_channel_spreading_gen}, and \eqref{trans-symbol}, we then reobtain the
system channel relation \eqref{gen.io.rel_4D}, with the channel coefficients $H_{l,k;l'\!,k'}$ expressed as
\begin{align}
  &H_{l,k;l'\!,k'} \eq \nonumber\\
   & \hspace*{3mm} e^{-\jmath 2 \pi \frac{N}{K}k'(l' \!-l)} \sum_{m=0}^{D-1} \sum_{i=0}^{J-1} T_h[m,i] \rmv
  \Bigg[\isum{n} \!\! \gamma^{*}[n] \ist e^{\jmath 2 \pi n \frac{k'-k}{K}} \nonumber\\
   & \hspace*{10mm} \times g[n\!-\!m\!-\!(l'\!-\!l)N] \ist \ppsi_{i}[n\!+\!lN] \Bigg]
     e^{-\jmath 2 \pi \frac{k'm}{K}}\rmv .
\label{H_T}
\end{align}
Note that the limiting cases $D \!=\! K$ and $J \!=\! N_{r}$ are also allowed.

\subsection{Compressive Channel Estimator}\label{sec:cs-chann-est-BEM}

The proposed compressive channel estimator operates in an iterative, decision-directed fashion. At the first iteration,
it utilizes the knowledge of some pilots $p_{l,k} \!\in\! \mathcal{A}$ with $(l,k) \!\in\! \mathcal{P}$.
The pilot position set $\mathcal{P}$ is selected uniformly at random within $\{0,\ldots,L\!-\!1\} \times \{0,\ldots,K\!-\!1\}$.
At later iterations, the estimator additionally uses {\it virtual} pilots, which are based on the symbol decisions produced by a suitable ISI/ICI equalizer (e.g.,
\cite{Schniter_sp04,rugini_CL05,Das_Schniter2007msi,Tauboeck2007,Hampejs2009})
followed by the quantizer. Typically, the equalizer will use the (estimated) channel coefficients $H_{l,k;l'\!,k'}$ only within a
certain ``off-diagonal bandwidth,'' i.e., for $|l-l'|\leq l_{\max}$ and $|k-k'|\leq k_{\max}$ (modulo $K$).

At the $r\ist$th iteration, let $p^{(r)}_{l,k}$ denote ``extended
pilots'' (pilots augmented by virtual pilots) on an \emph{extended
pilot position set} $\mathcal{P}^{(r)}$. This set is defined as
$\mathcal{P}^{(r)} \rmv\triangleq
\mathcal{H}^{(r)}\oplus\mathcal{V} = \big\{(l,k)=(l_{1}\rmv+\rmv
l_{2},(k_{1}\rmv +\rmv k_{2})\!\!\!\mod\! K):(l_{1},k_{1}) \rmv
\in\rmv  \mathcal{H}^{(r)} \rmv, \,(l_{2},k_{2}) \rmv \in\rmv
\mathcal{V} \big\}$, where $\mathcal{V} \triangleq \{(l,k):
l=-l_{\max},\ldots,l_{\max};\, k=-k_{\max},\ldots,k_{\max} \}$ and
$\mathcal{H}^{(r)}\rmv$ will be specified later. Note that by this
construction for an extended pilot in $\mathcal{H}^{(r)}\rmv$, all
neighboring symbols (which yield the largest interference) are
also included in $\mathcal{P}^{(r)}\rmv$. Then, for $(l,k) \in
\mathcal{H}^{(r)}\rmv$, relation \eqref{gen.io.rel_4D} can be
written as \be \label{4D-rel_offs} r_{l,k} \eq\!\! \sum_{(l'\!,k')
\,\in\, \{(l,k)\} \ist\oplus\ist \mathcal{V}} \!\!
H_{l,k;l'\!,k'}\, p_{l'\!,k'}^{(r)} \ist+\ist z^{(r)}_{l,k} \,,
\quad (l,k) \in \mathcal{H}^{(r)}, \ee where the
noise/interference term $z^{(r)}_{l,k}$ includes noise, ISI/ICI
from \emph{outside} the set $\{(l,k)\}\oplus\mathcal{V}$,
and---possibly---some additional errors if $p^{(r)}_{l'\!,k'} \neq
a_{l'\!,k'}$. If $\mathcal{V}$ is chosen sufficiently large, the
ISI/ICI part in $z^{(r)}_{l,k}$ is negligible. Inserting
\eqref{H_T} into \eqref{4D-rel_offs} yields the noisy 2-D
expansion \be \label{4D-rel_offs_a}
  r_{l,k} \eq\rmv \sum_{m=0}^{D-1}\sum_{i=0}^{J-1} \theta_{m,i} \, w^{(r)}_{m,i}[l,k ] +
  z^{(r)}_{l,k}\,, \quad (l,k) \in \mathcal{H}^{(r)},
\ee
with $\theta_{m,i} \!\triangleq\! T_h[m,i]$ and
$w_{m,i}^{(r)}[l,k]
\!\triangleq\! \sum_{(l'\!,k') \,\in\, \{(l,k)\} \ist\oplus\ist \mathcal{V}}\, p_{l'\!,k'}^{(r)}$\linebreak 
$\times\ist e^{-\jmath 2 \pi N k'(l' \!-l)/K} \big[\isum{n} \gamma^*[n] \ist e^{\jmath 2 \pi n (k'-k)/K}
  g[n\!-\!m\rmv- (l'\!\rmv-\rmv l)N] \ist \ppsi_{i}[n\rmv+\rmv lN] \big]\ist e^{-\jmath 2 \pi k'm/K}$.
Differently from \eqref{BEM_dft} and \eqref{BEM}, this is an
expansion of the demodulated symbols $r_{l,k}$ and not of the channel coefficients
$H_{l,k}$. Note also that the basis functions $w_{m,i}^{(r)}[l,k
]$ depend on the extended pilots $p_{l,k}^{(r)}$, $(l,k) \!\in\!
\mathcal{P}^{(r)}\rmv$.

Using a stacking as in Section \ref{sec:chest}, the expansion
\eqref{4D-rel_offs_a} can be expressed as $\mathbf{r}^{(r)} =
\mathbf{W}^{(r)} \ttheta + \mathbf{z}^{(r)}\rmv$, where the
$|\mathcal{H}^{(r)}|$-dimensional vectors $\mathbf{r}^{(r)}\rmv$
and $\mathbf{z}^{(r)}\rmv$, the $JD$-dimensional vector $\ttheta$,
and the $|\mathcal{H}^{(r)}| \times JD$ matrix
$\mathbf{W}^{(r)}\rmv$ are defined in an analogous manner as,
respectively, $\mathbf{h}^{({\rm p})}\rmv$, $\mathbf{z}$,
$\aalpha$, and $\mathbf{U}^{({\rm p})}\rmv$ in Section
\ref{sec:chest}. With $\mathbf{y}^{(r)} \triangleq
\mathbf{r}^{(r)}\rmv$, $\mathbf{\Phi}^{(r)} \triangleq
\mathbf{W}^{(r)} \mathbf{D}^{(r)}\rmv$, and $\mathbf{x}^{(r)}
\triangleq (\mathbf{D}^{(r)})^{-1} \ttheta$, where the diagonal
matrix $\mathbf{D}^{(r)}\rmv$ is chosen such that all columns of
$\mathbf{\Phi}^{(r)}\rmv$ have unit $\ell_2$-norm, we
obtain\footnote{The computation of the measurement matrix
essentially requires $L(2l_{\max}+1)(2k_{\max}+1) J$ FFTs of
length $K$. Note that $J$ is typically very small, cf. Section
\ref{sec:spars_bf}.} the \emph{measurement equation} (cf.\
\eqref{chest_sparse_reconstr}) $\mathbf{y}^{(r)} =
\mathbf{\Phi}^{(r)} \mathbf{x}^{(r)} + \mathbf{z}^{(r)}\rmv$. As
in Section \ref{sec:chest}, we would like to recover the
length-$JD$ vector $\mathbf{x}^{(r)}\rmv$ from the known
length-$|\mathcal{H}^{(r)}|$ vector $\mathbf{y}^{(r)}\rmv$. If the
basis functions $\ppsi_{i}[n]$ in \eqref{h_expans} and
\eqref{spread_fct_gen} are chosen such that $\T_{h}[m,i]$ (or,
equivalently, $\ttheta$) is sparse, then also $\mathbf{x}^{(r)} =
(\mathbf{D}^{(r)})^{-1} \ttheta$ is sparse. Hence, our problem is
again a sparse reconstruction problem of the form
\eqref{sparse_reconstr}, with dimensions $M \rmv=\rmv \text{dim}
\{ \mathbf{x}^{(r)} \} \rmv= JD$ and $Q \rmv=\rmv \text{dim} \{
\mathbf{y}^{(r)} \} \rmv=\rmv |\mathcal{H}^{(r)}|$. We can thus
use a CS recovery technique\footnote{Whether
$\mathbf{\Phi}^{(r)}\rmv$ satisfies the RIP with a small
restricted isometry constant depends on the basis functions
$\ppsi_{i}[n]$ as well as on the extended pilot position set
$\mathcal{P}^{(r)}$; hence, performance guarantees cannot be made
in general.} to obtain an estimate $\hat{\mathbf{x}}^{(r)}\rmv$ of
$\mathbf{x}^{(r)}\rmv$ and, in turn, an estimate
$\hat{\ttheta}^{(r)}\! = \mathbf{D}^{(r)}
\hat{\mathbf{x}}^{(r)}\rmv$ or, equivalently,
$\hat{\T}_{h}^{(r)}[m,i]$.

From $\hat{\T}^{(r)}_{h}[m,i]$, estimates of the channel
coefficients $H_{l,k;l'\!,k'}$ for all $l,l' \rmv
=0,\ldots,L\!-\!1$ and $k,k'\rmv =0,\ldots,$ $K\!-\!1$ are
obtained via \eqref{H_T}. Then, an ISI/ICI equalizer yields symbol
estimates $\tilde{a}_{l,k}^{(r)}\rmv$ and, subsequently, a
quantizer produces detected symbols $\hat{a}_{l,k}^{(r)}$, $l =
0,\ldots,L\!-\!1$, $k = 0,\ldots,K\!-\!1$. On $\mathcal{P}$, these
are replaced by the known pilots, i.e., we set
$\hat{a}_{l,k}^{(r)} \triangleq p_{l,k}$ for $(l,k) \!\in\!
\mathcal{P}$.

Next, we determine $\mathcal{H}^{(r+1)}\rmv$ as the largest subset
of $\{0,\ldots,L\!-\!1\}\times \{0,\ldots,K \!-\!1\}$ such that
the new extended pilot set $\mathcal{P}^{(r+1)}\triangleq
\mathcal{H}^{(r+1)}\oplus\mathcal{V}$ contains only ``reliable''
detected symbols $\hat{a}_{l,k}^{(r)}$, and we define the new
extended pilots as $p_{l,k}^{(r+1)}\triangleq\hat{a}_{l,k}^{(r)}$
for $(l,k) \in \mathcal{P}^{(r+1)}$. Here, following
\cite{Hampejs2009}, a detected symbol $\hat{a}_{l,k}^{(r)}$ will
be considered as ``reliable'' either if $(l,k) \!\in\!
\mathcal{P}$ or, for $(l,k) \!\notin\! \mathcal{P}$, if the
corresponding symbol estimate $\tilde{a}_{l,k}^{(r)}$ (result of
equalization, before quantization) is significantly closer to
$\hat{a}_{l,k}^{(r)}$ than to any other symbol in $\mathcal{A}$.
For example, for the QPSK alphabet
$\mathcal{A}\triangleq\{1\!+\!j,1\!-\!j,-1\!+\!j,-1\!-\!j \}$,
$\hat{a}_{l,k}^{(r)}$ will be considered as reliable either if
$(l,k) \!\in\! \mathcal{P}$ or if both
$\big|\Re\big\{\tilde{a}_{l,k}^{(r)} \big\}\big| > \epsilon$ and
$\big|\Im\big\{\tilde{a}_{l,k}^{(r)} \big\}\big| > \epsilon$ for a
certain threshold $\epsilon \rmv>\rmv 0$.

Proceeding iteratively in this fashion, we successively construct
extended pilots $p_{l,k}^{(r)}\rmv$, which are used to estimate
$T_h[m,i]$ and, via \eqref{H_T}, the channel coefficients
$H_{l,k;l'\!,k'}$. The reliability criterion ensures that most of
the extended pilots equal the true transmitted symbols. Since the
$p_{l,k}^{(r)}\rmv$ are improved with the iterations, we expect
$|\mathcal{H}^{(r+1)}| > |\mathcal{H}^{(r)}|$ in general. The
iterative algorithm is initialized with $p^{(0)}_{l,k} = p_{l,k}$
and $\mathcal{P}^{(0)} = \mathcal{H}^{(0)} = \mathcal{P}$ (for
$r\!=\!0$, $\mathcal{V}=\{0 \}$, whereas later $\mathcal{V} =
\{(l,k): l=-l_{\max},\ldots,l_{\max};\,
k=-k_{\max},\ldots,k_{\max} \}$). Accordingly, we use the
conventional one-tap equalizer (without ISI/ICI equalization) at
the first iteration. The algorithm is terminated either if the
difference between $\hat{H}^{(r+1)}_{l,k;l'\!,k'}$ and
$\hat{H}^{(r)}_{l,k;l'\!,k'}$ (measured by a suitable norm) falls
below a certain threshold or after a fixed number of iterations.
While a proof of convergence for this iterative algorithm is not
available, we always observed convergence for reasonably chosen
$\ppsi_{i}[n]$ (see Section \ref{sec:spars_bf}), $|\mathcal{P}|$,
and $\epsilon$.

The proposed algorithm is not limited to strongly dispersive channels. For weakly
dispersive channels, we simply set $\mathcal{V}=\{0 \}$ at all iterations and replace the ISI/ICI equalizer by the conventional
one-tap equalizer. This effectively amounts to a decision-directed, iterative extension of the compressive channel estimator discussed in
Sections \ref{sec:cs_chest}--\ref{sec:basis_opt}. This extension can improve the estimation accuracy. Moreover, it can
increase the spectral efficiency of the system even further, since the pilot set $\mathcal{P}$
can be chosen quite small due to the successive improvements achieved by the iterations.
However, these gains come at the cost of some additional complexity.

\subsection{Sparsity-Inducing Basis Functions}\label{sec:spars_bf}

The basis functions $\ppsi_{i}[n]$, $i=0,\ldots, N_{r}\!-\!1$ have to be chosen such that the generalized spreading function $\T_{h}[m,i]$
in \eqref{spread_fct_gen} is sparse. In particular, (\ref{dominant_impresp_cont}) suggests that
$\T_{h}[m,i]$ should be sparse for the single-scatterer channel $h^{(\tau_1,\nu_1)}(t,\tau) =
\delta(\tau\!-\!\tau_1) \, e^{\jmath 2\pi \nu_1 t }\rmv$, for all $\tau_1 \in [0, \tau_{\max}]$ and $\nu_1 \in [-\nu_{\max},\nu_{\max}]$.
For this channel,
\begin{align}
  \T_{h}[m,i] &\eq \phi^{(\nu_{1})}\Big(m - \frac{\tau_{1}}{T_{\rmv{\rm s}}}\Big) \, \vartheta^{(\nu_1)}[i] \,, \nonumber\\
   & \hspace*{3mm} \text{with} \;\;
  \vartheta^{(\nu)}[i] \,\triangleq \sum_{n=0}^{N_{r}-1}\! e^{\jmath 2\pi \nu n T_{\rmv{\rm s}}}  \, \ppsi^{*}_{i}[n] \,.
\label{vartheta_sparse}
\end{align}
The factor $\phi^{(\nu_{1})}\big(m - \tau_{1}/T_{\rmv{\rm s}}\big)$ (see \eqref{phi_def}) is already sparse due to its fast decay
as discussed in Section \ref{sec:sparsity}. Thus, we have to design the
$\ppsi_{i}[n]$ such that the factor $\vartheta^{(\nu)}[i]$ is sparse for all $\nu \in [-\nu_{\max},\nu_{\max}]$.

For this purpose, we can adapt the basis optimization of Section \ref{sec:basis_opt}. Let $\mathcal{D}
\ist\triangleq \big\{\nu_{\Delta}  d,\,\, d=-\lceil\nu_{\max}/\nu_{\Delta}\rceil,\dots, \lceil \nu_{\max}/\nu_{\Delta}\rceil \big\}$
with $\nu_{\Delta}=1/(2 \ist T_{\rmv{\rm s}} N_{r})$
and rewrite the second equation in \eqref{vartheta_sparse} as $\vvartheta^{(\nu)}= \mathbf{P} \ist \mathbf{e} ^{(\nu)}\rmv$, with the
length-$N_{r}$ vectors $\vvartheta^{(\nu)} \triangleq \big[ \vartheta^{(\nu)}[0] \,\cdots\, \vartheta^{(\nu)}[N_{r}\!-\!1] \big]^{T}\rmv$
and $\mathbf{e}^{(\nu)} \triangleq \big[1 \;\, e^{\jmath 2\pi \nu T_{\rmv{\rm s}}} \,\cdots\, e^{\jmath 2\pi \nu (N_{r}-1) T_{\rmv{\rm s}}} \big]^{T}\rmv$
and the unitary $N_{r} \times N_{r}$ matrix $\mathbf{P}$ with entries
${(\mathbf{P})}_{i+1,n+1} \rmv= \ppsi^{*}_{i}[n]$. Optimal basis functions $\ppsi_{i}[n]$ are now defined as
$\hat{\mathbf{P}} = \arg \min_{\mathbf{P} \in\, \mathcal{U}} \ist \sum_{\nu\in \mathcal{D}} {\|\mathbf{P} \ist \mathbf{e}^{(\nu)} \|}_{1}$,
so that the iterative optimization algorithm of Section \ref{sec:opt_algo} can be used.
However, for large $N_{r} \approx NL$, the computational cost of this approach is quite high.

As a practical alternative, we propose a construction of the $\ppsi_{i}[n]$ that involves discrete prolate spheroidal sequences (DPSSs) \cite{Sle78}.
Basis expansion models using DPSSs have been considered previously \cite{zemen_sp05}.
If their design parameters are chosen according to maximum Doppler frequency $\nu_{\max}$, sampling period $T_{\rmv{\rm s}}$, and blocklength
$N_{r}$, the corresponding functions $\vartheta^{(\nu)}_{\text{p}}[i]$ in \eqref{vartheta_sparse} will
have an effective support $\{0,\ldots,J\!-\!1\}$ for all $\nu \in [-\nu_{\max},\nu_{\max}]$, where $J$ is small compared with $N_{r}$.
Unfortunately, within this support interval, the $\vartheta^{(\nu)}_{\text{p}}[i]$ are not sparse in general.

We will therefore use a specific combination of DPSSs and DFT
basis functions, which yields functions $\vartheta^{(\nu)}[i]$
that are still effectively zero outside $\{0,\ldots,J\!-\!1\}$
but, within that interval, preserve the sparsity obtained with the
DFT basis. Let $\ppsi^{(\text{p})}_{i}[n]$, $n \in \Z$,
$i=0,\ldots,N_{r}\!-\!1$ denote the DPSSs that are bandlimited to
$[-\nu_ {\max}T_{\rmv{\rm s}},\nu_{\max} T_{\rmv{\rm s}}]$ and
have maximum energy concentration in $\{0,\ldots,N_{r}\!-\!1\}$
\cite{Sle78}. In what follows, the DPSSs
$\ppsi^{(\text{p})}_{i}[n]$ will be truncated to
$\{0,\ldots,N_{r}\!-\!1\}$. Then, for large $N_{r}$, the support
of $\vartheta^{(\nu)}_{\text{p}}[i] \triangleq
\sum_{n=0}^{N_{r}-1}\! e^{\jmath 2\pi \nu n T_{\rmv{\rm s}}}
\,\ppsi^{(\text{p})*}_{i}[n]$ is effectively contained in
$\{0,\ldots,J\!-\!1\}$ for all $\nu \in [-\nu_{\max},\nu_{\max}]$,
where $J \triangleq 2 J_{0} \rmv+\rmv J_{1}$ with $J_{0}\!
\triangleq \! \sleb$ and $J_{1}\!\geq \! 2$ a small integer. In
addition, we consider the $2J_{0} \! + \! 1$ orthonormal DFT basis
functions $\ppsi^{(\text{f})}_{i}[n] \triangleq (1/\sqrt{N_{r}})
\, e^{\jmath 2 \pi i n/N_{r}}$, $n=0,\ldots,N_{r}\!-\!1$, for
$i=-J_{0},\ldots,J_{0}$. For these $i$, $\nu[i]\triangleq i
/(N_{r}T_{\rmv{\rm s}})$ is in $[-\nu_{\max},\nu_{\max}]$. We thus
have for all $i_{1}=-J_{0},\ldots,J_{0}$ and
$i_{2}=J,\ldots,N_{r}\!-\!1$
\begin{align}
  \langle \ppsi^{(\text{f})}_{i_{1}}, \ppsi^{(\text{p})}_{i_{2}}\rangle
      & \eq \frac{1}{\sqrt{N_{r}}} \! \sum_{n=0}^{N_{r}-1} e^{\jmath 2 \pi \frac{i_{1} n}{N_{r}}} \, \ppsi^{ (\text{p})* }_{i_{2}}[n] \nonumber\\
   & \eq \frac{1}{\sqrt{N_{r}}} \, \vartheta^{(\nu[ i_{1}] )}_{\text{p}}[i_{2}] \nonumber\\
   &\,\approx\, 0 \,,
\label{orth_fourier_prolate}
\end{align}
because $\nu[i_1] \in [-\nu_{\max},\nu_{\max}]$ but $i_2 \not\in
\{0,\ldots,J\!-\!1\}$. That is, $\ppsi^{(\text{f})}_{i_{1}}$ and
$\ppsi^{(\text{p})}_{i_{2}}$ are effectively orthogonal for the
specified ranges of $i_1$ and $i_2$. Let us now define the
following ordered set of (in total $N_{r}$) DFT functions and
(truncated) DPSSs:
\begin{align*}
  &\mathcal{M}'\,\triangleq\, \big\{ \ppsi^{(\text{f})}_{-J_{0}} \ist,\ldots, \ist\ppsi^{(\text{f})}_{J_{0}}\ist,\ist \ppsi^{(\text{p})}_{2J_{0}+1} \ist, \ldots,\ist \ppsi^{(\text{p})}_{N_{r}-1}\big\} \,.
\end{align*}
\begin{figure*}[t]
\centering
\hspace*{0mm}\includegraphics[height=6cm,width=12cm]{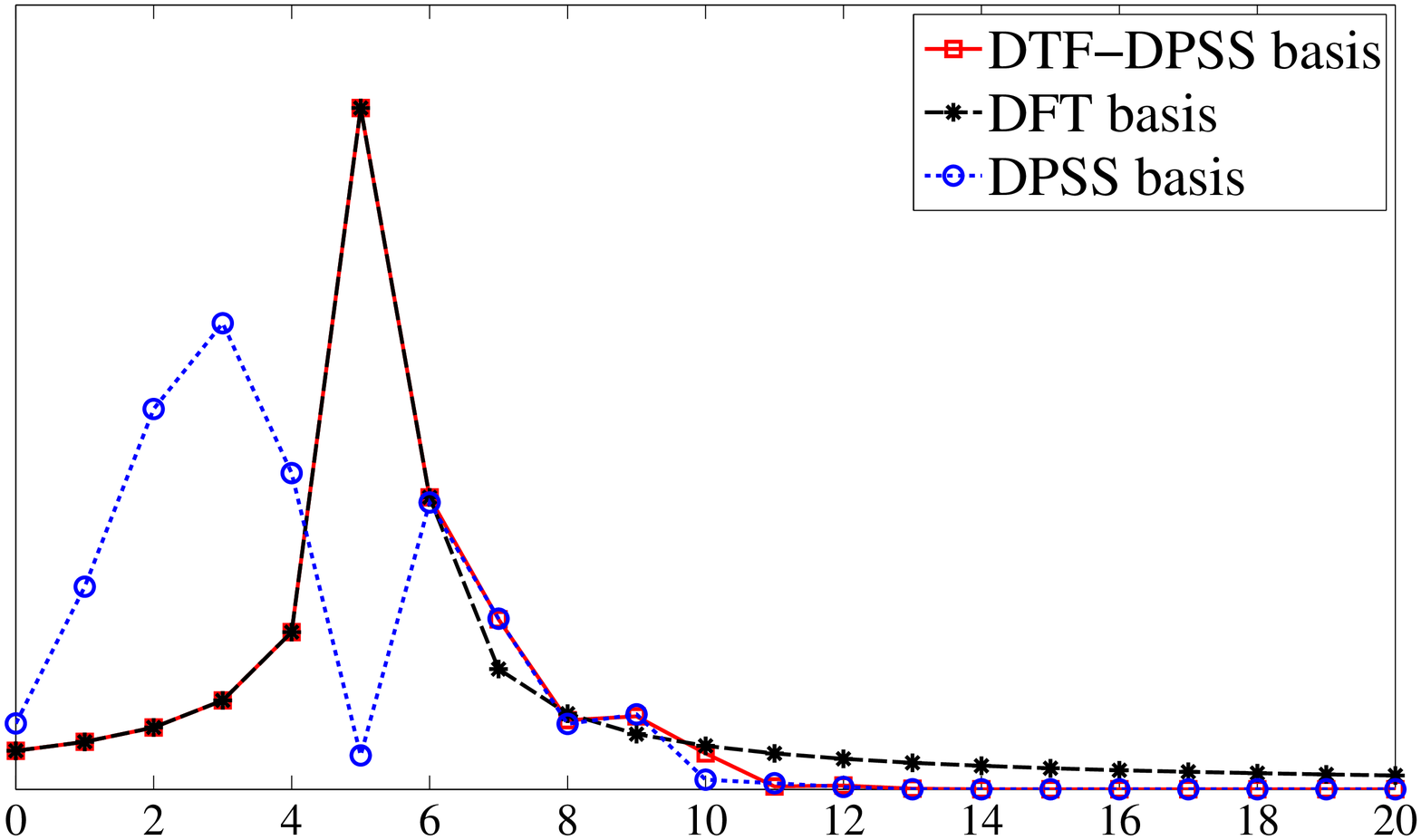}
\put(-166,-4){$i$} \vspace*{1mm}
\renewcommand{\baselinestretch}{1.1}\small\normalsize
\caption{Sparsity enhancement in $\vartheta^{(\nu)}[i]$ obtained
with the proposed combined DFT-DPSS basis, relative to a pure DFT
basis and a pure DPSS basis.} \label{fig.sparsity_mixed}
\vspace{-5mm}
\end{figure*}
Due to \eqref{orth_fourier_prolate} and the orthonormality of the
$\ppsi^{(\text{p})}_{i}$ \cite{Sle78}, all functions in
$\mathcal{M}'$ are (effectively) mutually orthonormal with the
exception of the DPSSs $\ppsi^{(\text{p})}_{i}$ within the index
range $i=2J_{0}\!+\!1,\ldots,J\!-\!1$, which are not orthonormal
to the DFT functions. Therefore, we derive the final set of basis
functions $\mathcal{M} \triangleq \{\ppsi_0, \ldots,
\ppsi_{N_{r}-1}\}$ by Gram-Schmidt orthonormalization
\cite{golub96} of $\mathcal{M}'\!$. This amounts to setting
$\ppsi_{i}=\ppsi^{(\text{f})}_{i-J_{0}}$ for $i\!=\!0,\ldots,2
J_{0}$ and $\ppsi_{i}= \sum_{n=0}^{2 J_{0}}c_n \ist
\ppsi^{(\text{f})}_{i-J_{0}} + \sum_{n=2J_{0} + 1}^{i}c_n \ist
\ppsi^{(\text{p})}_{n}$ for $i\!\geq\!2J_{0} \!+ \!1$, with
suitable coefficients $c_n$. It follows that $\langle
\ppsi_{i_{1}},\ppsi^{(\text{p})}_{i_{2}}\rangle\approx\, 0$ for
all $i_{1}=0,\ldots,J\!-\!1$ and $i_{2}=J,\ldots,N_{r}\!-\!1$.
Hence, the Gram-Schmidt orthonormalization algorithm yields
$\ppsi_{i}\approx\ppsi^{(\text{p})}_{i}$ for all $i
=J,\ldots,N_{r}\!-\!1$, i.e., the last $N_{r}\!-\!J$ basis
functions of $\mathcal{M}$ are effectively known \emph{a priori},
and the algorithm can therefore be terminated after $J$ steps. In
fact, only $J_{1}-1$ steps are required, because the first $2
J_{0} + 1=J-J_{1}+1$ (DFT) basis functions are also
known.

With this construction of the $\ppsi_{i}[n]$, the support of
$\vartheta^{(\nu)}[i] = \sum_{n=0}^{N_{r}-1}\! e^{\jmath 2\pi \nu
n T_{\rmv{\rm s}}} \,\ppsi^{*}_{i}[n]$ is approximately contained
in $\{0,\ldots,J\!-\!1\}$ for all $\nu \in [-\nu_
{\max},\nu_{\max}]$. Furthermore, for
$i\!=\!0,\ldots,J\!-\!J_{1}$, the $\ppsi_{i}[n]$ are DFT basis
functions, so that the sparsity of $\vartheta^{(\nu)}[i]$
corresponds to the sparsity given by the DFT basis for these
indices $i$. For the $J_{1}\!-\!1$ remaining indices
$i=J\!-\!J_{1}\!+\!1,\ldots, J\!-\!1$ within the support interval,
we cannot expect any sparsity of $\vartheta^{(\nu)}[i]$. However,
$J_{1}$ is quite small, so that the overall sparsity of
$\vartheta^{(\nu)}[i]$ is not deteriorated significantly.

For $N_{r} = N L = (2048 \rmv+\rmv 512) \ist 16 = 40960$ and
$\nu_{\max}T_{\rmv{\rm s}} = 0.2/K = 0.2/2048$ (corresponding to a
maximum Doppler frequency of 20\% of the subcarrier spacing),
Fig.\ \ref{fig.sparsity_mixed} depicts $|\vartheta^{(\nu)}[i]|$,
$i=0,\ldots,20$ for $\nu T_{\rmv{\rm s}} = 0.115/K = 0.115/2048$.
For comparison, $|\vartheta^{(\nu)}_{\text{f}}[i]|$ (obtained with
a pure DFT basis) and $|\vartheta^{(\nu)}_{\text{p}}[i]|$
(obtained with a pure DPSS basis) are also shown. We see that the
proposed DFT-DPSS basis leads to the sparsest result: for the pure
DPSS basis, there is no sparsity within the support interval,
while for the pure DFT basis, the sparsity is impaired by a strong
leakage effect.

\section{Simulation Results}\label{sec:simus}

Next, we demonstrate the performance gains that can be achieved with our
sparsity-enhancing basis expansions and estimation of ISI/ICI
channel coefficients, relative to the basic compressive estimator.
We show results for three different recovery algorithms, namely,
Lasso (equivalent to BP denoising), OMP, and CoSaMP.

\subsection{Simulation Setup}\label{sec:sim_setup}

{\em MC system parameters.} We simulated CP-OFDM systems with $K\!\in\!\{512,1024,2048\}$ subcarriers and CP length ratio
$(N\!-\!K)/K=1/4$. The systems employed 4-QAM symbols with Gray
labeling, a rate-$1/2$ convolutional code, and $32\!\times\! 16$ row-column interleaving. The
interpolation/anti-aliasing filters $f_{1}(t)=f_{2}(t)$ were
chosen as root-raised-cosine filters with roll-off factor $\rho \!=\! 1/4$.

{\em Recovery method.} For Lasso, we used the corresponding MATLAB
function from the toolbox SPGL1 \cite{SPGL1}. The required
regularization parameters were found by trial and error. CoSaMP
requires a prior estimate of the sparsity of $\mathbf{x}$. In all
simulations of Section \ref{sec:sim_BE_det}, we used the fixed
sparsity estimate $S \!=\! 262$, which was determined via the
formula $S = \lceil Q/(2 \log M)\rceil$ suggested in
\cite{netrXX}, where we set $Q \equiv |\mathcal{P}| \!=\! 2048$.
(Note that in most scenarios where CoSaMP was applied, we actually
used $2048$ pilots.) The number of CoSaMP iterations was $15$. For
OMP, we also used the sparsity estimate $S \!=\! 262$ (and, hence,
$262$ iterations), except for the strongly dispersive scenario of
Section \ref{sec:strongly_disp}. Therefore, in Section
\ref{sec:sim_BE_det}, the vectors produced by OMP and CoSaMP were
exactly $S$-sparse with $S \!=\! 262$.

{\em Channel.} We simulated and estimated the channel
during blocks of $L$ transmitted OFDM symbols ($L$ will be specified
in the individual subsections). For a more realistic simulation, the channel contained
a diffuse part in addition to a sparse (specular) part, with 20\,dB less total power than for the sparse part.
The scattering function of the diffuse part was bricked-shaped
within a rectangular delay-Doppler region $\{0,\ldots,K/4\!-\!1\} \times [-\nu_{\max}T_{\rmv{\rm s}},\nu_{\max}T_{\rmv{\rm s}}]$.
The discrete-delay-Doppler spreading function $S_h[m,i]$ of the sparse part was computed from
(\ref{Spread_fct_dom}). We always assumed $P\!=\!20$ propagation paths with scatterer
delay-Doppler positions $(\tau_p / T_{\rmv{\rm s}}, \nu_{p}
T_{\rmv{\rm s}})$ chosen uniformly at random within (or within a
subset of, cf.\ Section \ref{sec:sim_BE_det}) $\{0,\ldots,K/4\!-\!1\}
\times [-\nu_{\max}T_{\rmv{\rm s}},\nu_{\max}T_{\rmv{\rm s}}]$ for
each block of $L$ OFDM symbols. The scatterer amplitudes $\eta_{p}$ were randomly drawn from zero-mean, complex Gaussian distributions with three different
variances (3 strong scatterers of equal mean power, 7 medium scatterers with 10\,dB less mean power, and
10 weak scatterers with 20\,dB less mean power). Furthermore, we added complex white Gaussian noise $z[n]$
whose variance was adjusted to achieve a prescribed receive signal-to-noise ratio (SNR)
defined as (cf.\ \eqref{io_channel_impresp})
$\sum_{n=0}^{N_{r}-1} \E \{ \{|r[n]-z[n]|^2 \}/\sum_{n=0}^{N_{r}-1}\E\{\{|z[n]|^2 \}$.
\begin{figure*}[t]
\centering
\hspace*{-10mm}\includegraphics[height=9.5cm,width=20cm]{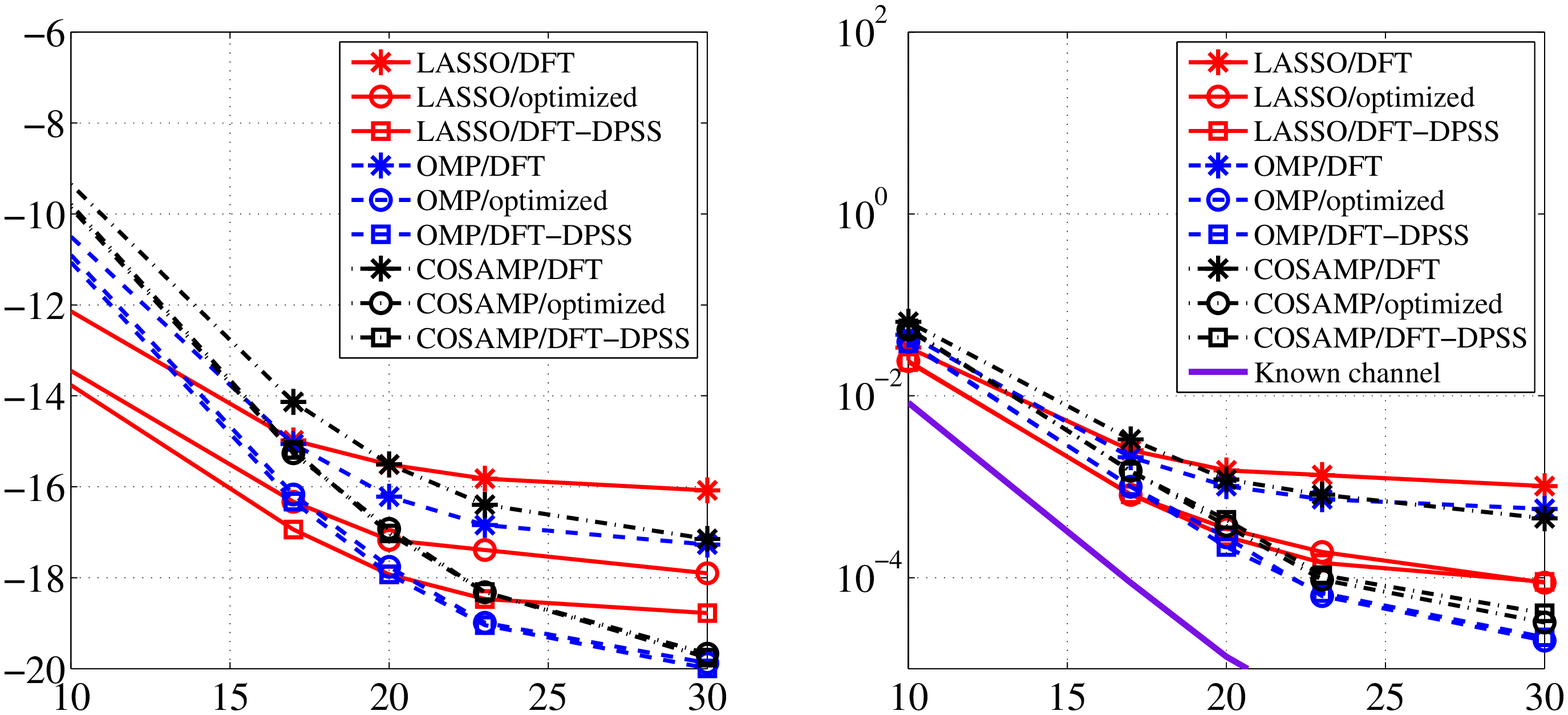}
\put(-407,10){(a)} \put(-155,10){(b)} \put(-426,27){ SNR [dB]}
\put(-170,27){SNR [dB]} \put(-533,117){ \begin{sideways}MSE
[dB]\end{sideways}} \put(-281,130){
\begin{sideways}BER\end{sideways}}
    \vspace*{-4mm}
\renewcommand{\baselinestretch}{1.1}\small\normalsize
\caption{Performance of compressive estimators versus the SNR: (a)
MSE, (b) BER.} \label{fig.SIM_SNR}
\end{figure*}

{\em Subsampling and pilots.} All estimators employed a subsampled time-frequency grid with $\Delta K \!=\! 4$ and $\Delta L \!=\!  1$,
on which the pilots were selected uniformly at random.

{\em Performance measures.} For all simulations, the performance
is measured by the mean square error (MSE) normalized by the mean
energy of the channel coefficients, as well as by the bit error
rate (BER).

\subsection{Performance Gains Through Basis Expansions}\label{sec:sim_BE_det}

We first compare the performance of compressive channel estimation
using the DFT basis (underlying the basic estimator of Section
\ref{sec:cs_chest}), the optimized basis of Section
\ref{sec:basis_opt} (without knowledge of channel statistics), and
the combined DFT-DPSS basis of Section \ref{sec:disp_channel}. The
number of subcarriers is $K \rmv=\rmv 2048$, the blocklength is $L
\rmv=\rmv 16$, and the maximum Doppler frequency is $\nu_{\max}
T_{\rmv{\rm s}} = 0.03/K$ (i.e., $3\%$ of the subcarrier spacing).
Here, the maximum Doppler frequency is quite small; accordingly,
the estimator of Section \ref{sec:cs-chann-est-BEM} only performs
its initial iteration (where $\mathcal{V} \rmv=\rmv \{0\}$). All
estimators use the same constellation of $|\mathcal{P}| \rmv=\rmv
2048$ pilots, corresponding to 6.25\% of all symbols. Fig.\
\ref{fig.SIM_SNR} depicts the performance versus the SNR for the
three recovery algorithms employed. The performance of the
optimized basis and the combined DFT-DPSS basis is seen to be
similar and clearly superior to that of the pure DFT basis,
especially at high SNR. This performance gain is due to the better
sparsity achieved, and it is obtained even though the coherence of
the optimized basis ($\mu_\mathbf{V} \!=\! 2.237$) is greater than
that of the DFT basis ($\mu_\mathbf{U} \!=\! 1$) and the
measurement matrix for the combined DFT-DPSS basis is not
constructed from an (ideally) unitary matrix. The larger gap to
the known-channel BER performance observed in Fig.\
\ref{fig.SIM_SNR}(b) at high SNR occurs because (i) the number of
pilots is too small for the channel's sparsity, and (ii) the
OMP-based and CoSaMP-based estimators produce $S$-sparse signals
with $S=262$, which is too small for the channel's sparsity.

\begin{figure*}[t]
\centering
\hspace*{-10mm}\includegraphics[height=9.5cm,width=20cm]{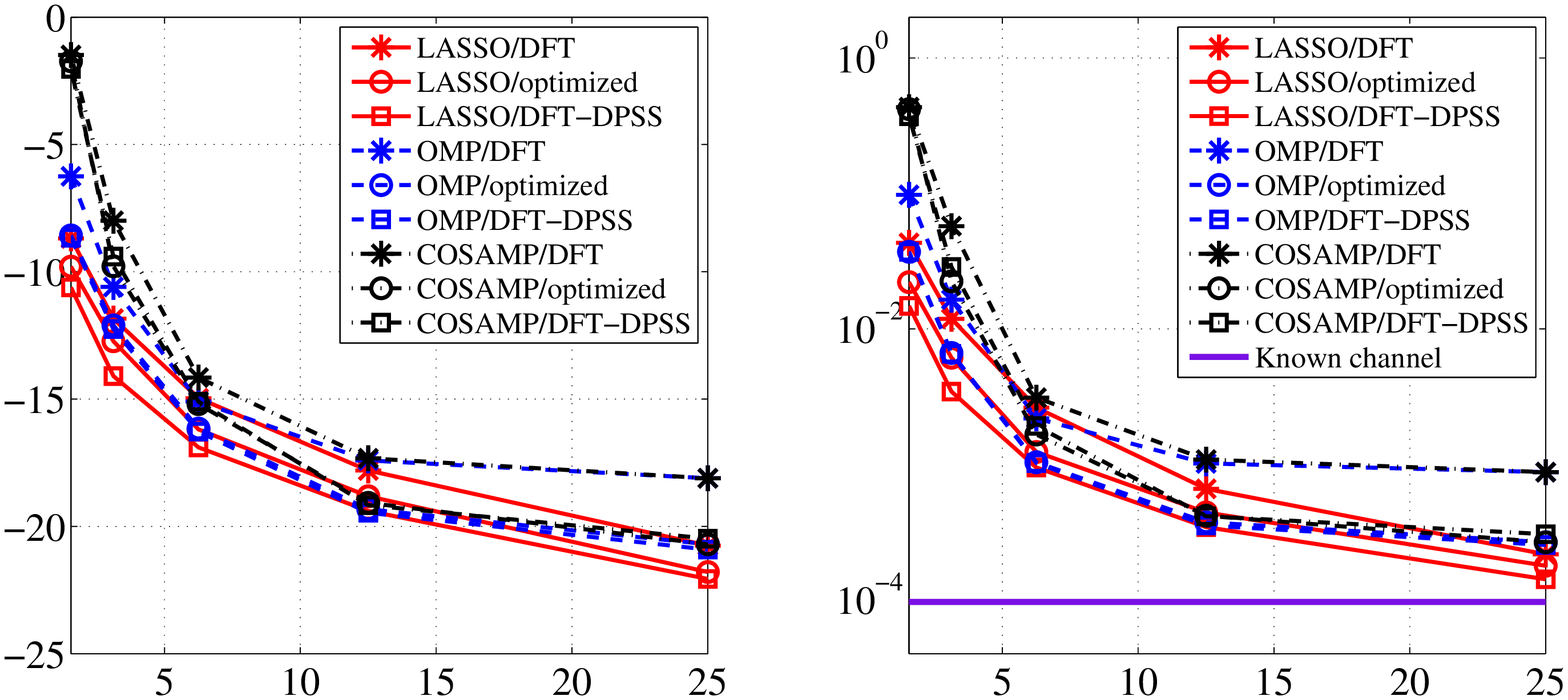}
\put(-407,10){(a)} \put(-155,10){(b)} \put(-419,27){$
|\mathcal{P}|$ [\%]} \put(-168,27){$|\mathcal{P}|$ [\%]}
\put(-533,117){ \begin{sideways}MSE [dB]\end{sideways}}
\put(-281,130){ \begin{sideways}BER\end{sideways}}
    \vspace*{-4mm}
\renewcommand{\baselinestretch}{1.1}\small\normalsize
\caption{Performance of compressive estimators versus the number
of pilots: (a) MSE, (b) BER.} \label{fig.SIM_pilots}
\end{figure*}

The number of pilots, $|\mathcal{P}|$, is an important design
parameter because it equals the number of measurements available
for sparse reconstruction. Fig.\ \ref{fig.SIM_pilots} depicts the
performance versus $|\mathcal{P}| \in \{512,\ldots,8192 \}$
(corresponding to 1.5625\% \!\dots 25\% of all symbols) at an SNR
of $17\,$dB. As a reference, the known-channel BER is also plotted
as a horizontal line. It is seen that, as expected, the
performance of all estimators improves with growing
$|\mathcal{P}|$. The optimized basis and the combined DFT-DPSS
basis are again superior to the DFT basis.

\begin{figure*}[t]
\vspace*{-13mm} \centering
\hspace*{-10mm}\includegraphics[height=9.5cm,width=20cm]{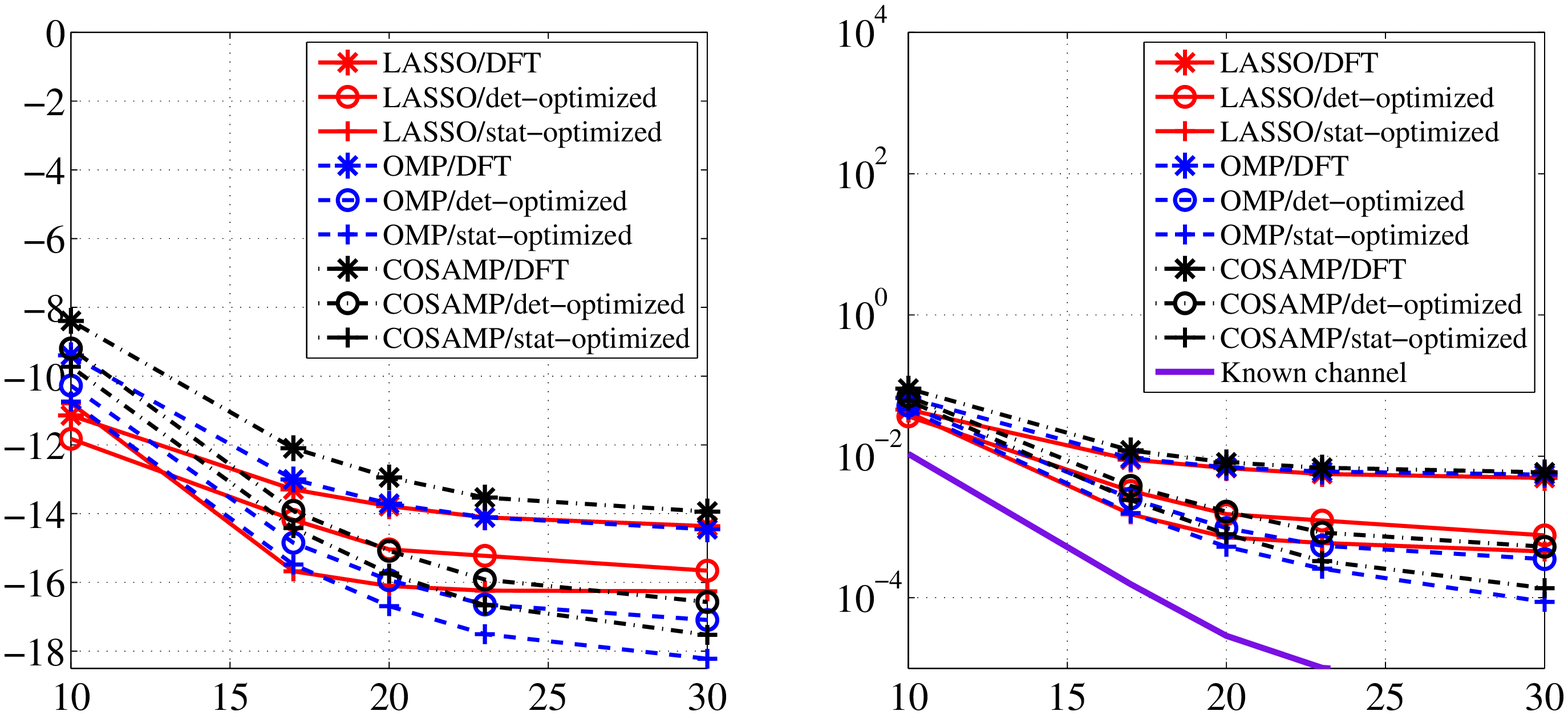}
\put(-407,10){(a)} \put(-155,10){(b)} \put(-426,27){ SNR [dB]}
\put(-170,27){SNR [dB]}
\put(-533,117){ \begin{sideways}MSE [dB]\end{sideways}}   
\put(-281,130){ \begin{sideways}BER\end{sideways}}
    \vspace*{-4mm}
\renewcommand{\baselinestretch}{1.1}\small\normalsize
\caption{Performance of DFT-based, deterministically optimized,
and statistically optimized compressive estimators versus the SNR:
(a) MSE, (b) BER.} \label{fig.SIM_stat_opt}
\end{figure*}

Next, we demonstrate performance gains that can be achieved by the statistically optimized basis expansion of Section
\ref{sec:basis_stat_opt}. The system and channel parameters are $K\!=\!512$, $L\!=\!64$, $\nu_{\max}
T_{\rmv{\rm s}} = 0.05/K$  ($5\%$ of the subcarrier spacing), and $|\mathcal{P}| \rmv=\rmv 2048$ (6.25\% of all symbols). For
the sparse channel part, the $20$ scatterer delay-Doppler
positions $(\tau_p / T_{\rmv{\rm s}}, \nu_{p} T_{\rmv{\rm s}})$
now are chosen uniformly at random only within $\{0,\ldots,127\} \times
([-0.05/K,-0.0375/K] \cup [0.0375/K,0.05/K])$. This serves as a rough approximation to the Jakes Doppler spectrum \cite{Jakes74},
according to which the scatterers are stronger when they are closer to the maximum Doppler frequency. In order to
optimize the basis expansion with this prior statistical
knowledge, the pdf $p(\tau_1,\nu_1)$ (see Section
\ref{sec:basis_stat_opt}) is set equal to a constant $c_1 \!>\!0$
within $[0,127 \, T_{\rmv{\rm s}}] \times ([-0.05/(K T_{\rmv{\rm
s}}),$ $-0.0375/(K T_{\rmv{\rm s}})] \cup [0.0375/(K T_{\rmv{\rm
s}}),0.05/(K T_{\rmv{\rm s}})])$ and equal to zero outside. The
variance of $\eta_{1}$ given $(\tau_1,\nu_1)$ is assumed constant,
i.e., $\sigma^2(\tau_1,\nu_1)=c_2 \!>\!0$. Fig.\
\ref{fig.SIM_stat_opt} depicts the resulting performance versus the SNR. For comparison, we also show the performance of
the deterministically optimized basis expansion, which uses only
knowledge of $\nu_{\max}$, as well as the performance of the DFT basis and the
known-channel BER performance. The statistically optimized basis is seen
to outperform the other bases. This can be explained by the
fact that it reduces the leakage effects occurring within the Doppler interval $[-0.0375/(K T_{\rmv{\rm s}}),0.0375/(K T_{\rmv{\rm s}})]$.

\subsection{Performance Gains Through ISI/ICI Coefficient Estimation}\label{sec:strongly_disp}

\begin{figure*}[t]
\centering
\hspace*{-10mm}\includegraphics[height=9.5cm,width=20cm]{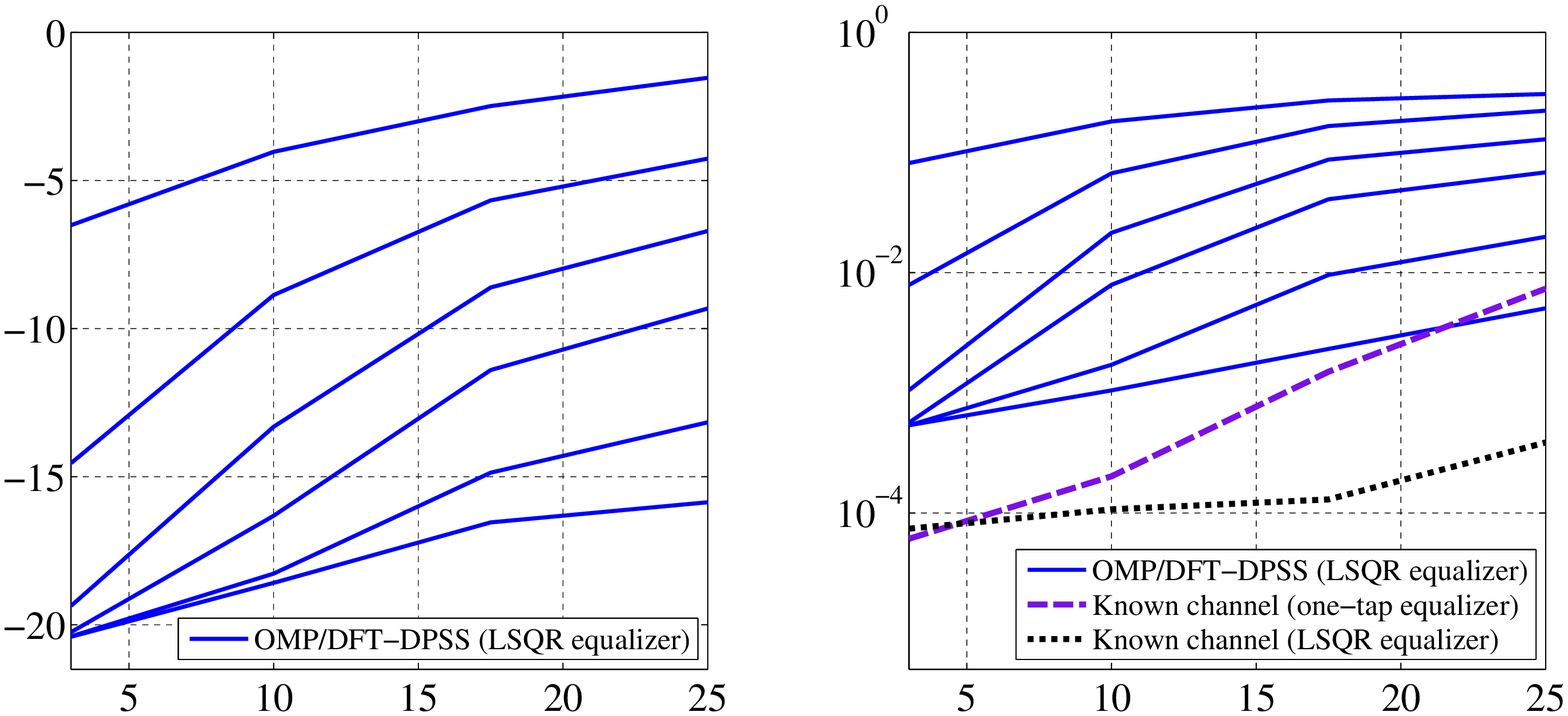}
\put(-407,10){(a)} \put(-155,10){(b)} \put(-433,27){ $\nu_{\max}
T_{\rmv{\rm s}} K$ [\%]} \put(-177,27){$\nu_{\max} T_{\rmv{\rm s}}
K$ [\%]} \put(-405,217.5){\small {\color{blue} $R\!=\!0$ (one-tap
equalizer)}} \put(-398,180){\small {\color{blue} $R\!=\!1$}}
\put(-397,155){\small {\color{blue} $R\!=\!2$}}
\put(-392,135){\small {\color{blue} $R\!=\!3$}}
\put(-378,111){\small {\color{blue} $R\!=\!5$}}
\put(-350,98){\small {\color{blue} $R\!=\!9$}}
\put(-200,213.5){\small {\color{blue} $R\!=\!0$ (one-tap
equalizer)}} \put(-192,193){\small {\color{blue} $R\!=\!1$}}
\put(-187,179){\small {\color{blue} $R\!=\!2$}}
\put(-183.5,166){\small {\color{blue} $R\!=\!3$}}
\put(-170,150.5){\small {\color{blue} $R\!=\!5$}}
\put(-150,140){\small {\color{blue} $R\!=\!9$}} \put(-533,117){
\begin{sideways}MSE [dB]\end{sideways}} \put(-281,130){
\begin{sideways}BER\end{sideways}}
    \vspace*{-4mm}
\renewcommand{\baselinestretch}{1.1}\small\normalsize
\caption{Performance of the decision-directed compressive
estimator versus the channel's maximum normalized Doppler
frequency for different numbers of iterations $R$: (a) MSE, (b)
BER.} \label{fig.SIM_doppler}
\end{figure*}
Finally, we assess the performance of the compressive, iterative,
decision-directed estimator of Section \ref{sec:disp_channel},
which is able to estimate also off-diagonal (ISI/ICI) channel
coefficients. We consider a wide range of maximum Doppler
frequencies, corresponding also to strongly frequency-dispersive
channels; more specifically, $\nu_{\max} T_{\rmv{\rm s}} \in
[0.03/K,0.25/K]$ or $3\% \ldots 25\%$ of the subcarrier spacing.
The system parameters are $K \!=\! 1024$, $L \!=\! 4$, $\text{SNR}
\!=\! 17\,$dB, and $|\mathcal{P}|=128$ (i.e., only 3.125\% of all
symbols). There occurs no ISI, only ICI. The estimator uses
$\mathcal{V}=\{(0,-3),\ldots,(0,3) \}$ for all iterations $r\geq
1$, so that the ICI equalizer processes the main diagonal plus the
first three upper and lower off-diagonals. The reliability
threshold is $\epsilon \rmv=\rmv 0.2$. For ICI equalization, we
use the LSQR equalizer proposed in \cite{Tauboeck2007}, with a
fixed number of $15$ iterations. Furthermore, we use OMP with $90$
iterations for CS recovery, and the combined DFT-DPSS basis of
Section \ref{sec:spars_bf}.

Fig.\ \ref{fig.SIM_doppler} depicts the performance of the
estimator versus the maximum Doppler frequency for iterations up
to $r\!=\!R$, with $R \!\in\! \{0,\ldots,9\}$. For comparison, the
known-channel BER performance of conventional one-tap equalization
and of LSQR-based ICI equalization is also shown. The MSE takes
into account the estimated diagonal and first three upper and
lower off-diagonal channel coefficients; it is normalized
accordingly. For $R = 0$, where only the diagonal channel
coefficients are estimated, the off-diagonal coefficients of the
estimated channel are set to zero when calculating the MSE. It is
seen from Fig.\ \ref{fig.SIM_doppler} that for $R = 0$, the
performance is very poor even for small $\nu_{\max}$ (weakly
dispersive channels). This is due to the small number of pilots
used. However, the performance is improved with an increasing
number $R$ of iterations, thus demonstrating the benefits of
off-diagonal coefficient estimation and the use of virtual pilots.
The initial improvement is slower for larger $\nu_{\max}$, again
because of the small number of pilots. It is furthermore seen that
for $R = 9$ iterations, for large $\nu_{\max}$, the proposed
compressive estimator is superior to the known-channel performance
of one-tap equalization. Our results also show that the proposed
decision-directed method is advantageous not only for coping with
strongly dispersive channels; it is equally useful for further
improving the spectral efficiency, even for mildly dispersive
channels, because of the smaller number of pilots required.

\section{Conclusion}\label{sec:conc}

We considered the application of compressed sensing techniques to
the estimation of doubly selective multipath channels within
pulse-shaping multicarrier systems (which include OFDM systems as
a special case). The channel coefficients on a subsampled
time-frequency grid are estimated in a way that exploits the
channel's sparsity in a dual delay-Doppler domain. We demonstrated
that this delay-Doppler sparsity is limited by leakage effects.
For combating leakage effects and, thus, enhancing sparsity, we proposed the use
of an explicit basis expansion that replaces the Fourier transform
used in the basic compressive channel estimation method. We also
developed an iterative basis design algorithm, and we extended our basis design to the case where prior statistical information about the channel
is available.

For strongly time-frequency dispersive channels, we then presented
an alternative compressive channel estimator that is capable of
estimating the ``off-diagonal'' channel coefficients
characterizing intersymbol and intercarrier interference
(ISI/ICI). Sparsity of the channel representation was here achieved by a
basis expansion combining the advantages of Fourier (exponential)
and prolate spheroidal sequences.

Simulation results demonstrated considerable performance gains
achieved by the proposed sparsity-enhancing basis expansions and by explicit
estimation of ISI/ICI channel coefficients. The additional
computational complexity required by the basis expansions is
moderate; in particular, the bases can be precomputed before the
start of data transmission.

\section*{Acknowledgments}\label{sec.ack}

The authors would like to thank Prof.\ G.\ Matz and Dr.\ P.\ Fertl for helpful
discussions. They are also grateful to the anonymous reviewers for numerous constructive comments that have resulted in a major
improvement of this paper.


\renewcommand{\baselinestretch}{1.09}\small\normalsize\small

\bibliographystyle{ieeetr}

\begin{biography}[{\includegraphics[width=1in,height=1.25in,clip,keepaspectratio]{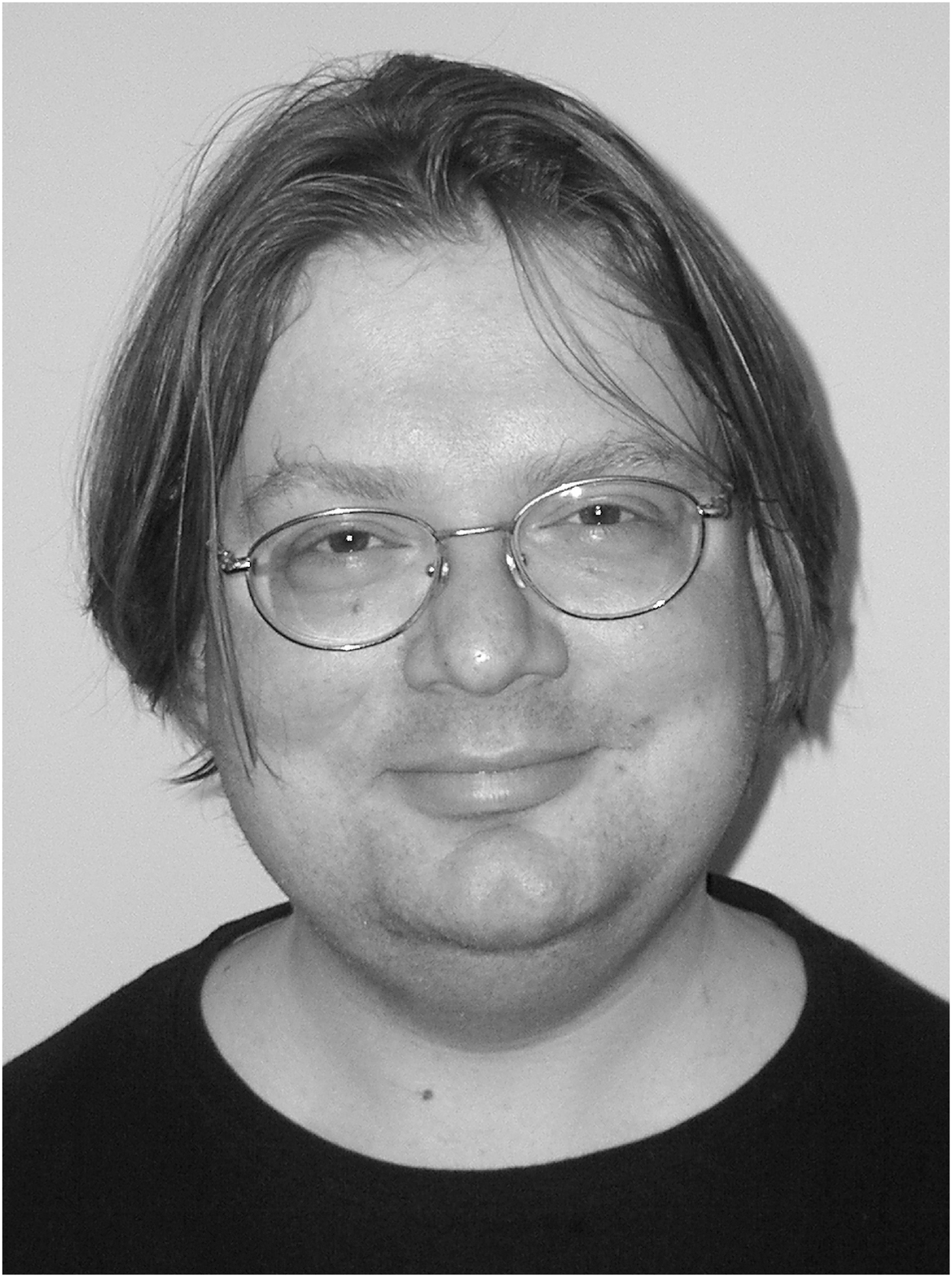}}]{Georg Taub\"{o}ck}
(S'01--M'07) received the Dipl.-Ing. degree and the Dr.techn.
degree (with highest honors) in electrical engineering and the
Dipl.-Ing. degree in mathematics (with highest honors) from Vienna
University of Technology, Vienna, Austria in 1999, 2005, and 2008,
respectively. He also received the diploma in violoncello from the
Conservatory of Vienna, Vienna, Austria, in 2000.

From 1999 to 2005, he was with the FTW Telecommunications Research
Center Vienna, Vienna, Austria, and since 2005, he has been with
the Institute of Communications and Radio-Frequency Engineering,
Vienna University of Technology, Vienna, Austria.

His research interests include wireline and wireless
communications, compressed sensing, signal processing, and
information theory.

\end{biography}

\begin{biography}[{\includegraphics[width=1in,height=1.25in,clip,keepaspectratio]{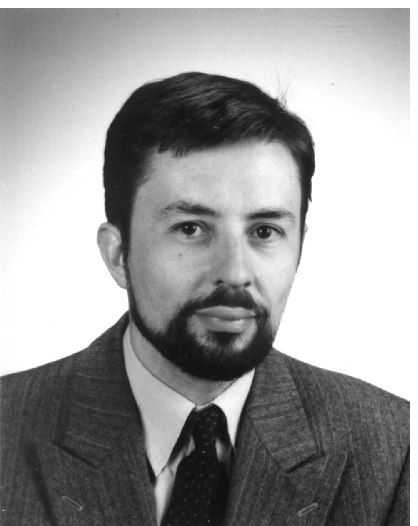}}]{Franz Hlawatsch}
(S'85--M'88--SM'00) received the Diplom-Ingenieur, Dr.\ techn., and
Univ.-Dozent (habilitation) degrees in electrical
engineering/signal processing from Vienna University of
Technology, Vienna, Austria in 1983, 1988, and 1996, respectively.

Since 1983, he has been with the Institute of Communications and
Radio-Frequency Engineering, Vienna University of Technology,
where he is currently an Associate Professor. During 1991--1992, as
a recipient of an Erwin Schr\"{o}dinger Fellowship, he spent a
sabbatical year with the Department of Electrical Engineering,
University of Rhode Island, Kingston, RI, USA. In 1999, 2000, and
2001, he held one-month Visiting Professor positions with
INP/ENSEEIHT/TeSA, Toulouse, France and IRCCyN, Nantes, France. He
(co)authored a book, a review paper that appeared in the IEEE
Signal Processing Magazine, about 180 refereed scientific papers
and book chapters, and three patents. He coedited two books. His
research interests include signal processing for wireless
communications, statistical signal processing, and compressive
signal processing.

Prof. Hlawatsch was Technical Program Co-Chair of EUSIPCO 2004 and
served on the technical committees of numerous IEEE conferences.
From 2003 to 2007, he served as an Associate Editor for the IEEE
TRANSACTIONS ON SIGNAL PROCESSING, and since 2008, he has served
as an Associate Editor for the IEEE TRANSACTIONS ON INFORMATION
THEORY. From 2004 to 2009, he was a member of the IEEE SPCOM
Technical Committee. He is coauthor of a paper that won an IEEE
Signal Processing Society Young Author Best Paper Award.

\end{biography}

\begin{biography}[{\includegraphics[width=1in,height=1.25in,clip,keepaspectratio]{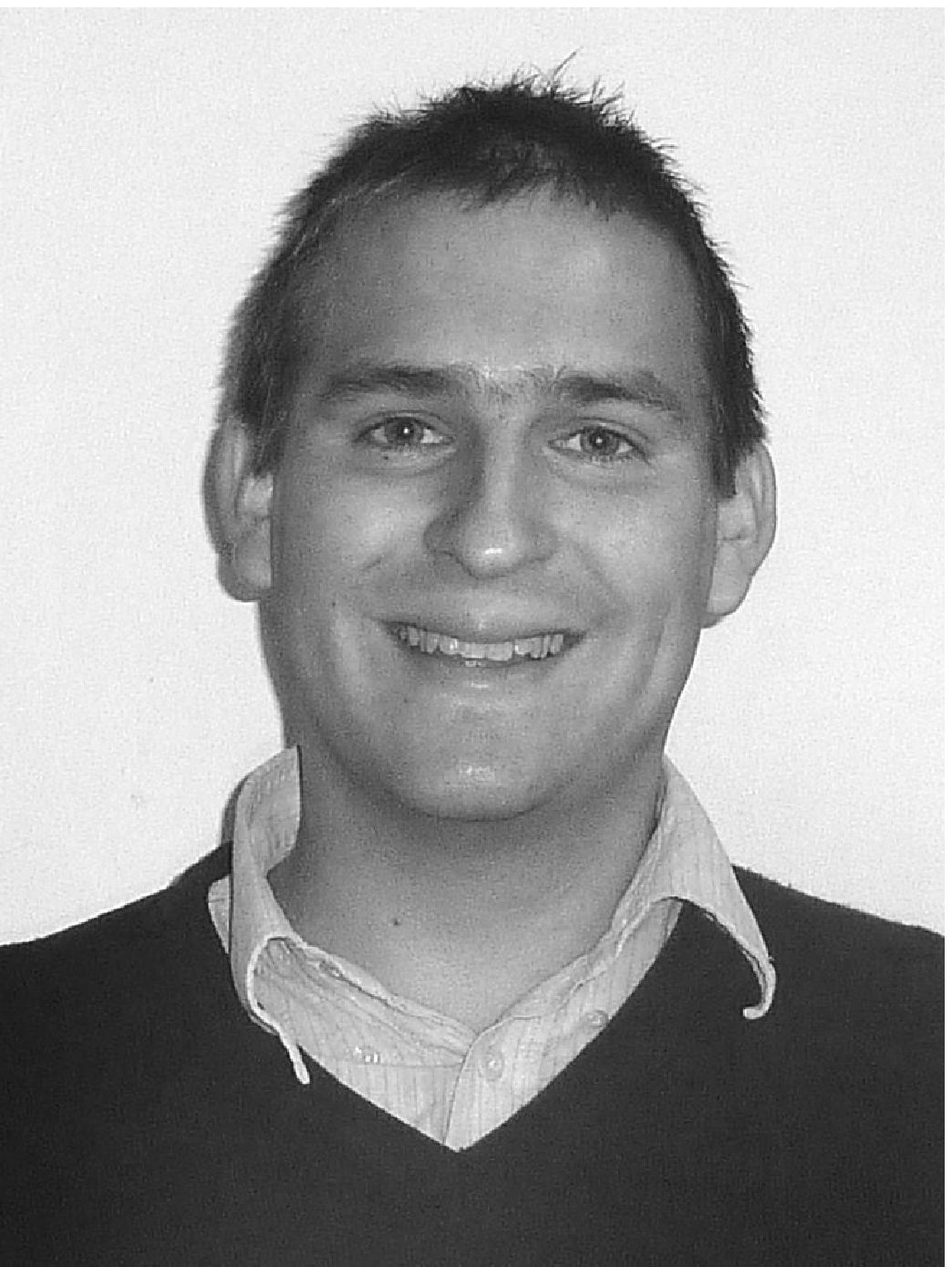}}]{Daniel Eiwen}
(S'10) received the diploma degree in mathematics from the
University of Vienna in 2008. Since September 2008, he has been
with the Numerical Harmonic Analysis Group (NuHAG) at the Faculty
of Mathematics, University of Vienna, where he pursues a PhD
degree.

His research interests include compressed sensing, sparse
approximation, and time-frequency analysis, as well as their
application in signal processing.

\end{biography}

\begin{biography}[{\includegraphics[width=1in,height=1.25in,clip,keepaspectratio]{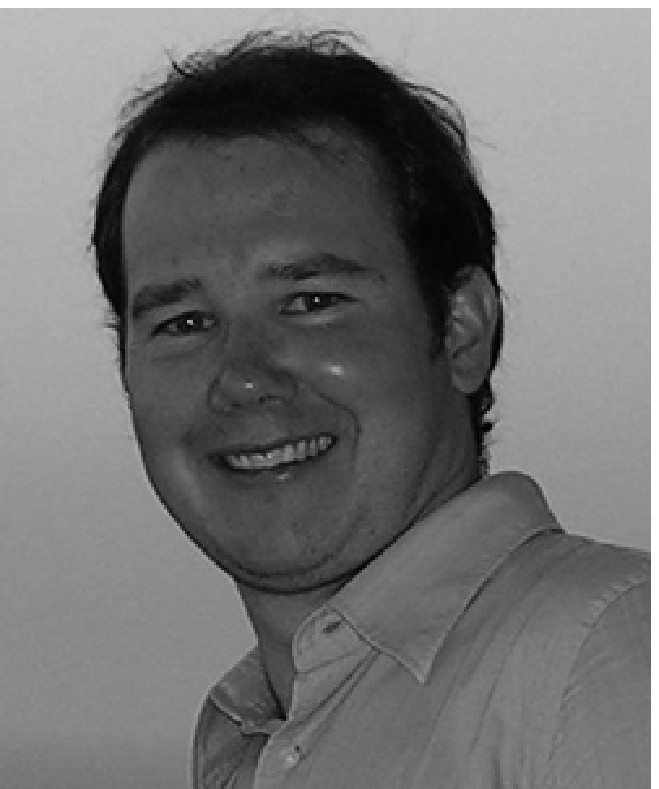}}]{Holger Rauhut}
received the diploma degree in mathematics from the Technical
University of Munich in 2001. He was a member of the graduate
program Applied Algorithmic Mathematics at the Technical
University of Munich from 2002 until 2004, and received the Dr.\
rer.\ nat.\ degree in mathematics in 2004. From 2005 until 2008, he
was with the Numerical Harmonic Analysis Group at the Faculty of
Mathematics, University of Vienna as a PostDoc. Since March 2008,
he has been a professor for mathematics (Bonn Junior Fellow) with
the Hausdorff Center for Mathematics and the Institute for
Numerical Simulation, University of Bonn, Germany.

His research interests include compressed sensing, sparse
approximation, random matrices, time-frequency and wavelet
analysis.

\end{biography}

\end{document}